\pdfoutput=1
\let\oldvec\vec
\documentclass[traditabstract]{aa}
\let\vec\oldvec

\usepackage{graphicx}
\usepackage{amssymb,amsmath}
\usepackage{txfonts}
\usepackage{natbib}
\usepackage{hyperref}
\usepackage{rotating}
\usepackage{color, colortbl}

\renewcommand{\vec}[1]{\mathbf{#1}}

\definecolor{LightCyan}{rgb}{0.88,1,1}

\definecolor{Gray}{gray}{0.9}

\DeclareSymbolFont{mymathvariables}{OT1}{cmr}{m}{n}
\SetSymbolFont{mymathvariables}{normal}{OT1}{cmr}{m}{n}
\DeclareSymbolFontAlphabet{\mathnormal}{mymathvariables}
\DeclareMathSymbol{v}{\mathalpha}{mymathvariables}{118}

\begin{document}

\title{Langmuir Wave Electric Fields Induced by Electron Beams in the Heliosphere}

   \author{Hamish A. S. Reid and Eduard P. Kontar}


   \institute{SUPA School of Physics and Astronomy, University of Glasgow, G12 8QQ, UK}

\abstract{Solar electron beams responsible for type III radio emission generate Langmuir waves as they propagate out from the Sun.  The Langmuir waves are observed via in-situ electric field measurements.  These Langmuir waves are not smoothly distributed but occur in discrete clumps, commonly attributed to the turbulent nature of the solar wind electron density.  Exactly how the density turbulence modulates the Langmuir wave electric fields is understood only qualitatively.  Using weak turbulence simulations, we investigate how solar wind density turbulence changes the probability distribution functions, mean value and variance of the beam-driven electric field distributions.  Simulations show rather complicated forms of the distribution that are dependent upon how the electric fields are sampled.  Generally the higher magnitude of density fluctuations reduce the mean and increase the variance of the distribution in a consistent manor to the predictions from resonance broadening by density fluctuations.  We also demonstrate how the properties of the electric field distribution should vary radially from the Sun to the Earth and provide a numerical prediction for the in-situ measurements of the upcoming Solar Orbiter and Solar Probe Plus spacecraft.}

\titlerunning{Beam-induced Electric Field Distributions}
\authorrunning{Reid and Kontar}

\maketitle

\section{Introduction} \label{sec:intro}

Solar type III radio bursts are believed to be caused by high-energy electron beams propagating through the corona or the solar wind.  The beams experience a beam-plasma instability that generates an enhanced level of Langmuir waves.  The Langmuir waves can then undergo wave-wave processes to emit electromagnetic radiation near the local plasma frequency and at the harmonic.  Since the introduction of this qualitative theory by \citet{Ginzburg:1958aa} there has been an intensive research effort to understand the radio emission mechanism from solar electron beams; motivated by the diagnostic potential to further understand particle acceleration and transport through the solar corona and the solar wind.

The theory underwent an initial dilemma \citep{Sturrock:1964aa} that for typical coronal and beam parameters, the Langmuir wave growth rate is so efficient a beam would lose almost all energy within a few metres of propagation.  To explain interplanetary type III bursts, electron beams must travel distances of 1~AU and beyond.  It was later suggested \citep{Zaitsev:1972aa} that a spatially limited electron beam could solve this dilemma by the back of the electron beam absorbing the Langmuir wave energy produced by the front of the electron beam. The continuous generation and re-absorption of Langmuir waves means their peak energy density effectively travels at the speed of the electron beam, despite the Langmuir wave group velocity being much lower.  Beam propagation with Langmuir waves over distances of 1~AU was subsequently simulated using quasilinear relaxation \citep{Takakura:1976aa,Magelssen:1977aa}.

The enhanced electric fields from Langmuir waves were first observed accompanying energetic electrons and type III radio emission at 1~AU using the IMP-6 and IMP-8 spacecraft \citep{Gurnett:1975aa} but it was the subsequent in-situ observations using the Helios spacecraft around 0.45~AU \citep{Gurnett:1976aa,Gurnett:1977aa} that solidified the theory of \citet{Ginzburg:1958aa}.  A wide range of electric field strengths were observed accompanying type III emission \citep[see][for a number of examples]{Gurnett:1978aa} with peak values ranging from 50~mV/m around 0.4~AU, down to 0.3 mV/m around 1~AU.  \citet{Gurnett:1980aa} fitted a power-law to 86 events occurring at various distances from the Sun and found a dependence of $r^{-1.4}$ for the peak electric field.  However, in all the observations the Langmuir waves were bursty (or clumpy) in nature, with the mean electric field strength noticeably below the peak values \citep[e.g.][]{Gurnett:1978aa,Robinson:1993aa}.  To describe the clumpy behaviour the original theory of \citet{Ginzburg:1958aa} required additional physics.

The bursty or clumpy behaviour of Langmuir waves is normally attributed to the small-scale density fluctuations in the background solar wind plasma \citep[e.g][]{Smith:1979aa,Melrose:1980aa,Muschietti:1985aa,Melrose:1986aa}.  The variation in background electron density $\Delta n_e$ refracts the Langmuir waves \citep{Ryutov:1969aa}, changing Langmuir wave $k$-vectors.  The Langmuir waves distribution is then modulated, dependent upon the efficiency of refraction by the background density fluctuations, and causes Langmuir waves to appear in clumps \citep[see e.g.][for recent studies]{Ratcliffe:2012aa,Bian:2014aa,Voshchepynets:2015aa}.  In-situ measurements of density turbulence in the solar wind at 1~AU estimate values of 
$\Delta n/n$ around $10-1$\%  over a wide range of length scales with a power density spectra that has different spectral indices at different scales \citep[e.g.][]{Celnikier:1983aa,Celnikier:1987aa,Malaspina:2010ab,Chen:2013ab}.  Density fluctuations parallel to the magnetic field might have a lower intensity as measurements suggest the magnetic field turbulence parallel to the magnetic field is smaller \citep[e.g.][]{Chen:2012aa} and the level of $\Delta n$ is correlated to $\Delta B$ \citep[e.g.][]{Howes:2012aa}.  \citet{Robinson:1992aa,Robinson:1993aa} argue that the beam propagates in a state close to marginal stability where wave generation (or beam relaxation) is balanced by the effects of density fluctuations and assume that the growth of waves becomes stochastic and normally distributed.  The probability distribution function (PDF) of the electric fields measured in-situ during a type III event have been analysed \citep[][]{Robinson:1993aa,Gurnett:1993aa,Vidojevic:2012aa}.  Above a background around 0.01~mV/m the PDF had a drop-off, fit by a power-law in \citet{Gurnett:1993aa} and a parabola in \citet{Robinson:1993aa}, with one PDF measured by \citet{Robinson:1993aa} having a characteristic field above the background at 0.035~mV/m.  Using background subtraction \citet{Vidojevic:2012aa} found the PDFs from 36 events were better approximated by a Pearson distribution, mostly of type I.  The Langmuir wave PDF has been measured in other environments \citep{Robinson:2004aa} including the Earth's foreshock where the PDF forms a power-law with a negative spectral index when averaged over large distances \citep{Cairns:1997aa,Bale:1997aa}, and can be fit with a log-normal distribution when analysed over shorter distances \citep{Sigsbee:2004aa}.

To understand the resonant interaction with propagating electrons and Langmuir waves a large number of numerical studies using quasilinear theory \citep{Vedenov:1963aa,Drummond:1964aa} have been undertaken \citep[see e.g.][and references therein]{Takakura:1976aa,Magelssen:1977aa,Grognard:1985aa,Kontar:2001ab,Foroutan:2007aa,Kontar:2009aa,Li:2014aa,Ratcliffe:2014aa,Ziebell:2015aa,Reid:2015aa}.  The simulations show that an electron beam was found to fully relax to a plateau in velocity space as it propagates through plasma with almost constant speed \citep{MelNik:1999ab}. However, 
the large-scale background density gradient from the radially decreasing solar corona and solar wind plasma density refracts waves to high $\vec{k}$-values, causing the electron beam to propagate with decreasing speed \citep{Kontar:2001ab}, likely to be responsible for the deceleration of type III sources \citep{Krupar:2015aa}.   This energy loss changes an initial power-law energy spectrum into a broken power-law in transit to 1~AU \citep{Kontar:2009aa,Reid:2013aa}.  However, the large-scale background density gradient does not cause a clumpy Langmuir wave distribution.  Using quasilinear theory, a subset of simulations have modelled the clumpy distribution of Langmuir waves induced from an electron beam with the inclusion of small-scale background density fluctuations \citep[e.g.][]{Kontar:2001aa,Reid:2010aa,Li:2012aa,Ratcliffe:2012aa,Voshchepynets:2015aa}.  Langmuir waves interacting with an electron beam over small-scales have also been analysed using the Zakharov Equations \citep[e.g.][]{Zaslavsky:2010aa,Krafft:2013aa,Krafft:2014ab}, the particle-in-cell approach \citep[e.g.][]{Pecseli:1992aa,Tsiklauri:2011aa,Karlicky:2012aa,Pecseli:2014aa} and Vlasov simulations \citep[e.g.][]{Umeda:2007aa,Henri:2010aa,Daldorff:2011aa}.  The approaches found electrons relax to a plateau in the velocity distribution and that density fluctuations can cause particles to be accelerated at the high velocity end of the beam.

We study here the modification of the electric field distribution produced during propagation by the electron beam cloud.  We systematically look at how the intensity of the density turbulence modifies the induced distribution of Langmuir waves (and associated electric fields) over length scales longer than typically considered in previous studies; required to capture the behaviour between the Sun and the Earth.  We first show in Section \ref{sec:theory} how the electric field will be distributed in a plasma without density fluctuations.  After describing the numerical modelling in Section \ref{sec:model} we analyse the propagation of an electron beam through a background plasma with a constant mean density in Section \ref{sec:Earth}, similar to propagation near the Earth.  We demonstrate the effect of density fluctuations on the probability distribution function (PDF) of the electric field.  We then consider an electron beam injected in the solar atmosphere and propagating through the decreasing density of the solar corona in Section \ref{sec:Sun} and the inner heliosphere in Section \ref{sec:helio}.  The latter is an effort to predict what Solar Orbiter and Solar Probe Plus will observe of their upcoming journey towards the Sun.

\section{Electric field distribution in uniform plasma} \label{sec:theory}

\begin{figure}\center
  \includegraphics[width=0.99\columnwidth]{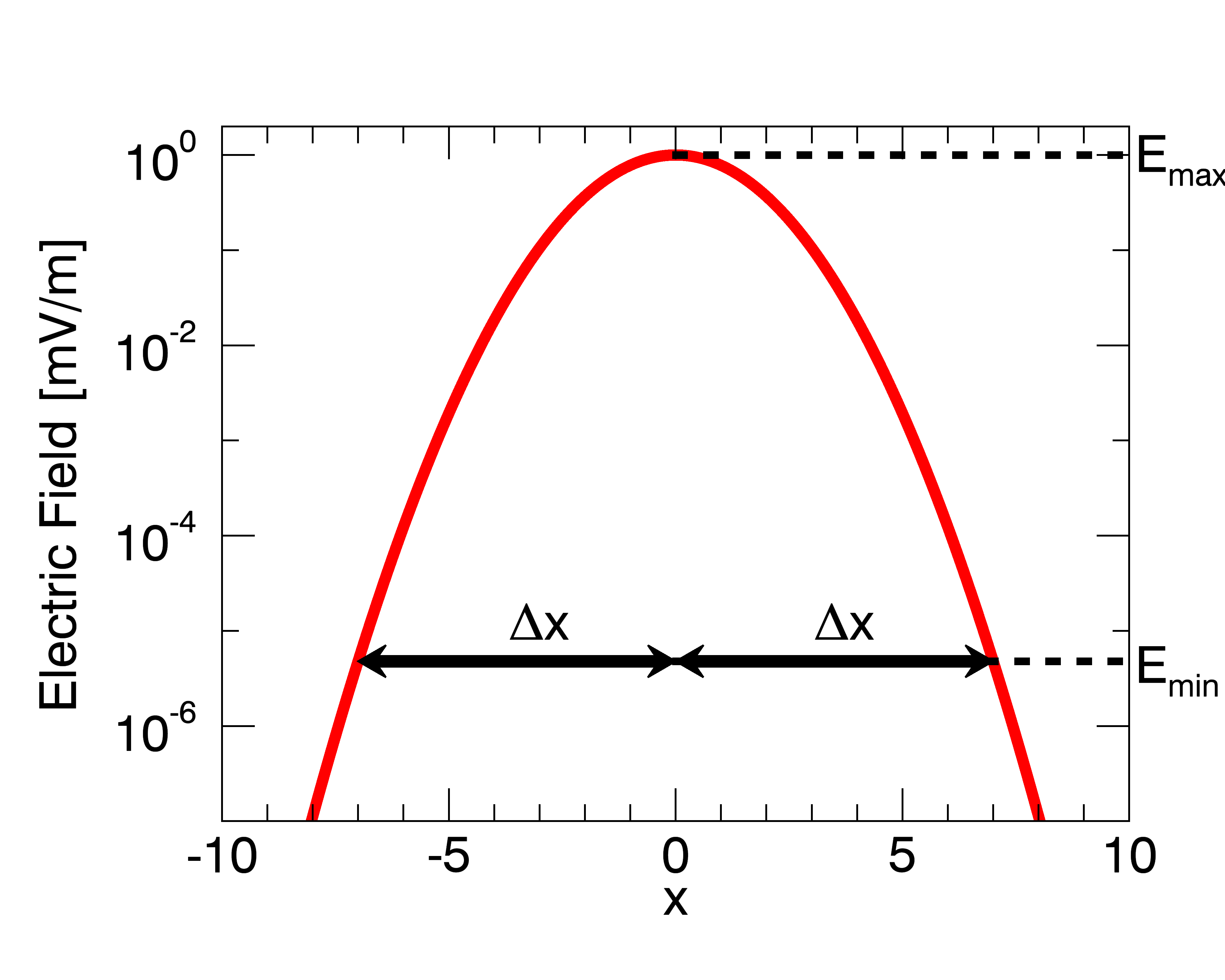}
  \includegraphics[width=0.99\columnwidth]{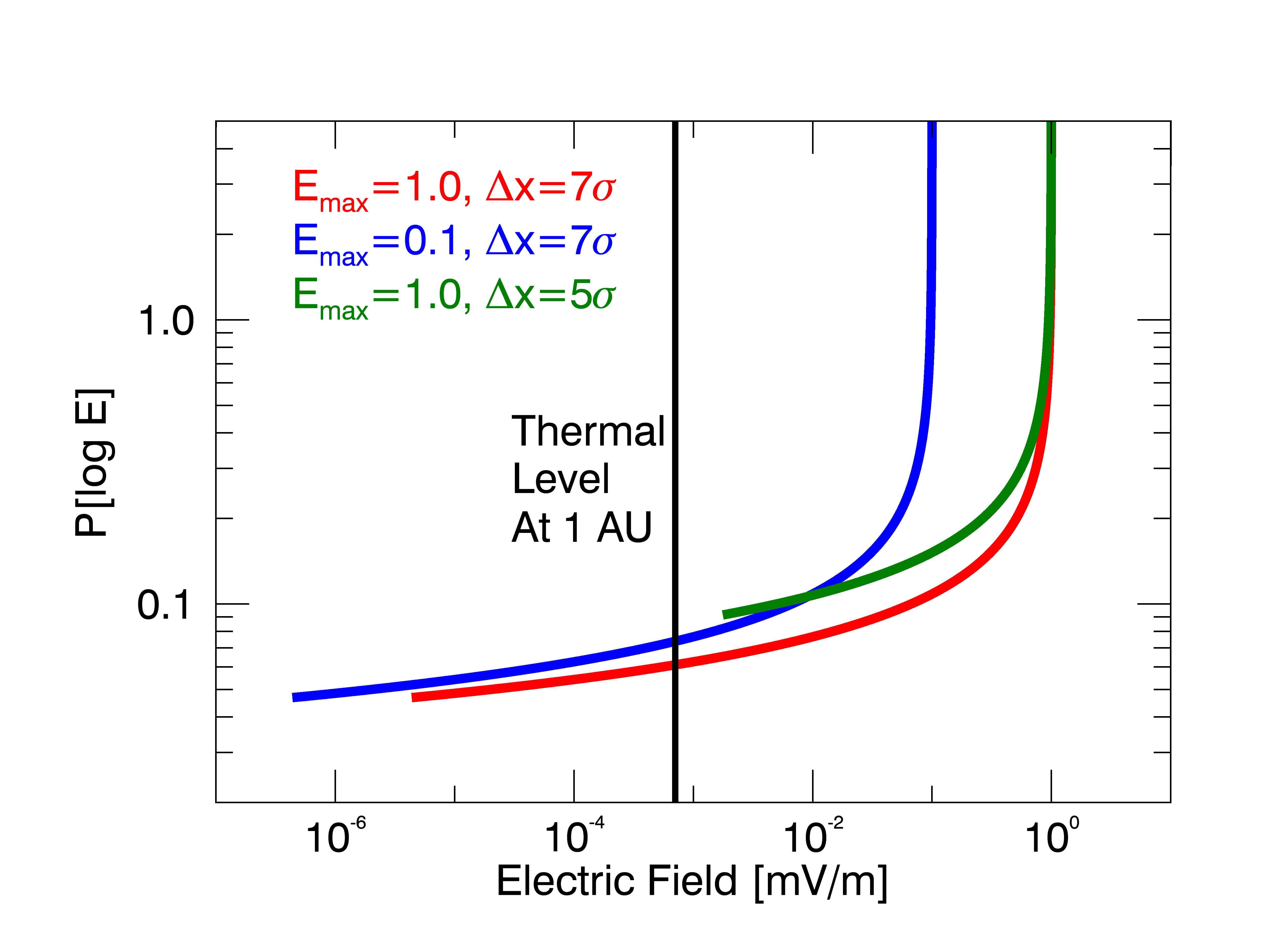}
\caption{Top: electric field described by Equation \ref{eqn:gauss} where $E_{max}=1$ and $\sigma=1$.  The sampling region $\Delta x=7\sigma$ is shown.  Bottom: probability density function (PDF) for $\log E$ over the region $\Delta x$.  The PDF is also shown when $E_{max}=0.1$~mV/m and $\Delta x=5\sigma$.  The black vertical line shows the thermal electric field at 1~AU when $n_e=5~\rm{cm}^{-3}$.}
\label{fig:Gauss_PDF}
\end{figure}

Observationally the electric field is measured in-situ as a function of time by a spacecraft drifting through plasma of the solar wind.  The electric field $E$ of the Langmuir waves is related to the Langmuir wave energy density $U_w=E^2/(8\pi)$.  If we approximate an electron beam by a Gaussian distribution in space along the direction of the magnetic field and assume uniform plasma, Langmuir waves have also a Gaussian distribution in space \citep[e.g.][]{MelNik:1999ab}.  Following gas-dynamic theory \citep{MelNik:1999ab}, the electric field induced by the Langmuir waves will take the form
\begin{equation}\label{eqn:gauss}
E^2(x)= 8\pi U_w(x) =E_{max}^2\exp\left(- \frac{x^2}{2\sigma^2}\right),
\end{equation}
where $\sigma$ is the characteristic spatial size of the electron cloud (and beam-driven Langmuir waves), $E_{max}$ is the maximum electric field.  The growth of Langmuir waves and hence the magnitude of the electric field occurs over many orders of magnitude and so it is appropriate to consider the logarithm of electric field (in this work we consider log in base 10).  For a Gaussian spatial distribution (Equation \ref{eqn:gauss}) the probability distribution function (PDF) of $\log E$ is proportional to the inverse square root of $\log E$.
\begin{equation}
P(\log E)\propto \log^{-0.5}\left(\frac{E_{max}}{E}\right).
\end{equation}
We select a sampling region that is symmetric around the peak ($-\Delta x,\Delta x$), where $\log E$ changes from $\log E_{min}\rightarrow\log E_{min}$ (see Figure \ref{fig:Gauss_PDF}) then $\Delta x=2\sigma\ln^{0.5}(E_{max}/E_{min})$.  Normalising the PDF to unity, $2\int_{\log E_{min}}^{\log E_{max}} P(\log E)d\log E=1$,
\begin{equation}\label{eqn:gauss_pdf}
P(\log E)=\frac{1}{4}\log^{-0.5}\left(\frac{E_{max}}{E_{min}}\right)\log^{-0.5}\left(\frac{E_{max}}{E(x)}\right).
\end{equation}
We have shown $P(\log E)$ in Figure \ref{fig:Gauss_PDF} where $E_{max}=1$~mV/m and $\Delta x=7\sigma$.  The thermal level of the electric field \citep{Meyer-Vernet:1979aa} is indicated for the plasma density $n_e=5~\rm{cm}^{-3}$, typical for the solar wind density near the Earth.  Figure \ref{fig:Gauss_PDF} also shows how the PDF changes for a different spatial distribution and sampling region $\Delta x$.  This emphasises that a change in the spatial profile (Equation \ref{eqn:gauss}) alters the shape of the PDF.  It also highlights the importance of sampling in the shape of the PDF.



The PDF of the electric field in non-uniform plasma of the solar wind will be a combination of the spatial distribution of electrons and the effect of density fluctuations that we explore in the following sections.

\section{Simulation Method} \label{sec:model}

\subsection{Model}

To investigate the effect of density fluctuations on the electric fields from Langmuir waves produced by a propagating electron beam we use self-consistent numerical simulations \citep{Kontar:2001ad}.  We model the time evolution of an electron beam through their distribution function $f(v,x,t)$ and the Langmuir waves through their spectral energy density $W(v,x,t)$.  The background plasma distribution function is assumed to be static in time.  The time evolution of the electrons and Langmuir waves are approximated using the following 1D kinetic equations
\begin{eqnarray}\label{eqk1}
\frac{\partial f}{\partial t} + \frac{v}{M(r)}\frac{\partial}{\partial r}M(r)f =
\frac{4\pi ^2e^2}{m_e^2}\frac{\partial }{\partial v}\left(\frac{W}{v}\frac{\partial f}{\partial v}\right)
\quad\quad\quad\quad\quad\cr
	 +\frac{4\pi n_e e^4}{m_e^2}\ln\Lambda\frac{\partial}{\partial v}\frac{f}{v^2} + S(v,r,t),
\end{eqnarray}
\begin{eqnarray}\label{eqk2}
\frac{\partial W}{\partial t} + \frac{\partial \omega_L}{\partial k}\frac{\partial W}{\partial r}
-\frac{\partial \omega _{pe}}{\partial r}\frac{\partial W}{\partial k}
= \frac{\pi \omega_{pe}}{n_e}v^2W\frac{\partial f}{\partial v}
 \quad\quad\quad\quad\cr
- (\gamma_{L} +\gamma_c )W + e^2\omega_{pe}v f \ln{\frac{v}{v_{Te}}},
\end{eqnarray}
where propagation is considered to be along a guiding magnetic field.  A complete description of Equations \ref{eqk1} and \ref{eqk2} can be found in our previous works \citep[e.g.][]{Reid:2015aa}.  Equation \ref{eqk1} simulates the electron propagation together with a decrease in density as the guiding magnetic flux rope expands (modelled through the cross-sectional area $M(r)$ of the expanding flux tube).  Equation \ref{eqk1} and \ref{eqk2} have the quasilinear terms \citep{Vedenov:1963aa,Drummond:1964aa} that describe the resonant wave growth ($\omega_{pe}=kv$) and the subsequent diffusion of electrons in velocity space .  The absorption of waves from the background plasma is modelled via the Landau damping rate $\gamma_L$.  Equations \ref{eqk1} and \ref{eqk2} also model the effect of electron and wave collisions with the background ions where $\ln\Lambda$ is the Coulomb logarithm and $\gamma_c$ is the collisional rate of Langmuir waves.  The collisional terms modify $f(v,x,t)$ and $W(v,x,t)$ primarily in the dense solar corona and have little effect in the rarefied plasma of the solar wind.  Equation \ref{eqk2} models the spontaneous emission of waves \citep[e.g.][]{Zheleznyakov:1970aa,Hannah:2009aa}, the propagation of waves, and importantly the refraction of waves on density fluctuations \citep[e.g.][]{Ryutov:1969aa,Kontar:2001aa}.

\subsection{Initial Electron beam}\label{sec:beam}

We use a source of electrons $S(v,r,t)$ in the form separating velocity, space and time
\begin{equation}\label{eqn:source}
S(v,r,t) = A_v v^{-\alpha}\exp\left(-\frac{r^2}{d^2}\right)A_t\exp\left(-\frac{(t-t_{inj})^2}{\tau^2}\right),
\end{equation}
where the velocity distribution is assumed to be a power-law characterised by $\alpha$ the velocity spectral index.  The constant $A_v\propto n_{beam}$ scales the injected distribution such that the integral over velocity between $v_{min}$ and $v_{max}$ gives the number density $n_{beam}$ of injected electrons.

The spatial distribution is characterised by $d$ [cm], the spread of the electron beam in distance.  The distance term is not normalised so increasing $d$ increases the number of electrons that are injected into the simulation and can be used with $n_{beam}$ to determine the total number of electrons injected into the simulation.

The temporal profile is characterised by $\tau$ [seconds] that governs the temporal injection profile.  The constant $A_t$ normalises the temporal injection such that the integral over time is 1.  The characteristic time $\tau$ does not control the number of injected electrons but does affect the injection rate.  The constant $t_{inj}=4\tau$.

\subsection{Background plasma}\label{sec:plasma}

Similar to previous works \citep[e.g.][]{Reid:2013aa}, the thermal level of Langmuir waves is set to
\begin{equation}\label{eqn:init_w}
W^{init}(v,r,t=0) = \frac{k_BT_e}{4\pi^2}\frac{\omega_{pe}^2}{v^2}\ln\left(\frac{v}{v_{Te}}\right),
\end{equation}
where $k_B$ is the Boltzmann constant and $T_e$ is the electron temperature.  Equation \ref{eqn:init_w} represents the thermal level of spontaneously emitted Langmuir waves from an uniform Maxwellian background plasma when Coulomb collisions are neglected. 

The mean background electron density $n_0(r)$ is constant for the simulations replicating conditions near the Earth.  This approximation is because at 1~AU the contribution of the density gradient from the solar wind expanding outwards from the Sun is small over the simulation distance.  For simulations where the electrons are injected at the Sun and propagate towards the Earth, we calculate $n_0(r)$ using the Parker model \citep{Parker:1958aa} to solve the equations for a stationary spherical symmetric solution \cite{Kontar:2001ab}, with a normalisation factor found from satellites \citep{Mann:1999aa}.  The density model is very similar to other solar wind density models like the Sittler-Guhathakurta model \citep{Sittler:1999aa} and the Leblanc model \citep{Leblanc:1998aa} except that the density is higher close to the Sun, below 10 $R_\odot$.  The density model reaches $5\times10^9~\rm{cm}^{-3}$ at the low corona and is more indicative to the flaring Sun, compared to $10^9~\rm{cm}^{-3}$ in the Newkirk model \citep{Newkirk:1961aa} or $10^8~\rm{cm}^{-3}$ in the Leblanc model.

For modelling the density fluctuations we first note that the power spectrum of density fluctuations near the Earth has been observed in-situ to obey a Kolmogorov-type power law with a spectral index of $-5/3$ \citep[e.g.][]{Celnikier:1983aa,Celnikier:1987aa,Chen:2013ab}.  Following the same approach of \citet{Reid:2010aa}, we model the spectrum of density fluctuations with a spectral index $-5/3$ between the wavelengths of $10^8$~cm and $10^{10}$~cm, so that the perturbed density profile is given by
\begin{equation}\label{eqn:fluc}
n_e(r) = n_0(r)\left[1 + C(r)\sum_{n=1}^N\lambda_n^{\mu/2}\sin(2\pi r/\lambda_n + \phi_n)\right]\,,
\end{equation}
 where $N=1000$ is the number of perturbations, $n_0(r)$ is the initial unperturbed density as defined above, $\lambda_n$ is the wavelength of the $n$-th fluctuation, $\mu=5/3$ is the power-law spectral index in the power spectrum, and $\phi_n$ is the random phase of the individual fluctuations.  $C(r)$ is the normalisation constant the defines the r.m.s. deviation of the density $\sqrt{\langle \Delta n(r)^2 \rangle}$ such that
 \begin{equation}
C(r) = \sqrt{\frac{2\langle \Delta n(r)^2 \rangle}{\langle n(r) \rangle^2\sum_{n=1}^N\lambda_n^{\mu}}}.
\end{equation}
Our one-dimensional approach means that we are only modelling fluctuations parallel to the magnetic field and not perpendicular.

Langmuir waves are treated in the WKB approximation such that wavelength is smaller than the characteristic size of the density fluctuations.  We ensure that the level of density inhomogeneity \citep{Coste:1975aa,Kontar:2001aa} satisfies
\begin{equation}
\frac{\Delta n}{n} < \frac{3k^2v_{th}^2}{\omega_{pe}^2}.
\end{equation}
The background fluctuations are static in time because the propagating electron beam is travelling much faster than any change in the background density.   To capture the statistics of what a spacecraft would observe over the course of an entire electron beam transit we look at the Langmuir wave energy distribution as a function of distance at one point in time.

\vspace{20pt}
\begin{center}
\begin{table*}
\centering
\begin{tabular}{ c  c  c  c  c  }

\hline\hline

Energy Limits & Velocity Limits & Spectral Index & Temporal Profile & Density Ratio \\ \hline
26~eV to 10~keV &  $4-80~v_{th}$ & $\alpha=4.0$& $\tau=20$~s & $n_b/n_e=2\times10^{-4}$ \\

\hline
\end{tabular}
\vspace{10pt}
\caption{Injected beam parameters for the electron beam travelling through plasma with a constant background density.}
\label{tab:beam_earth}
\end{table*}
\end{center}

\section{Electron beams near the Earth} \label{sec:Earth}

We explore the evolution of the electric field from the Langmuir waves induced by a propagating electron beam.  The beam is injected into plasma with a constant mean background electron density of $n_0=5~\rm{cm}^{-3}$ (plasma frequency of 20~kHz), similar to plasma parameters around 1~AU, near the Earth.  To explore how the intensity of density fluctuations influences the distribution of the induced electric field we add density fluctuations to the background plasma with varying levels of intensity.  The constant mean background electron density means that we know any changes on the distribution of the electric fields are caused by modifying the intensity of the density turbulence.  The background plasma temperature was set to $10^5$~K, indicative of the solar wind core temperature at 1~AU \citep[e.g.][]{Maksimovic:2005aa}, giving a thermal velocity of $\sqrt{k_b T_e / m_e}=1.2 \times 10^{8}~\rm{cm~s}^{-1}$.  The electron beam is injected into a simulation box that is just over 8 solar radii in length, representing a finite region in space around 1~AU.  To fully resolve the density fluctuations we used a spatial resolution of 200~km.

The beam parameters are given in Table \ref{tab:beam_earth}.  The energy limits are typical of electrons that arrive at 1~AU co-temporally with the detection of Langmuir waves \citep[e.g.][]{Lin:1981aa}.  The spectral index is obtained from the typical observed in-situ electron spectra below 10 keV near the Earth \citep{Krucker:2009aa}.  We note that this spectral index is lower than what is measured in-situ at energies above 50~keV, and inferred from X-ray observations \citep{Krucker:2007aa}.  The high characteristic time broadens the electron beam, a process that would have happened to a greater extent if our electron beam had travelled to 1~AU from the Sun.  The high density ratio is to ensure a high energy density of Langmuir waves is induced.

\subsection{Beam-induced electric field}

\begin{figure*}\center
  \includegraphics[width=0.99\columnwidth,trim=50 30 0 20,clip]{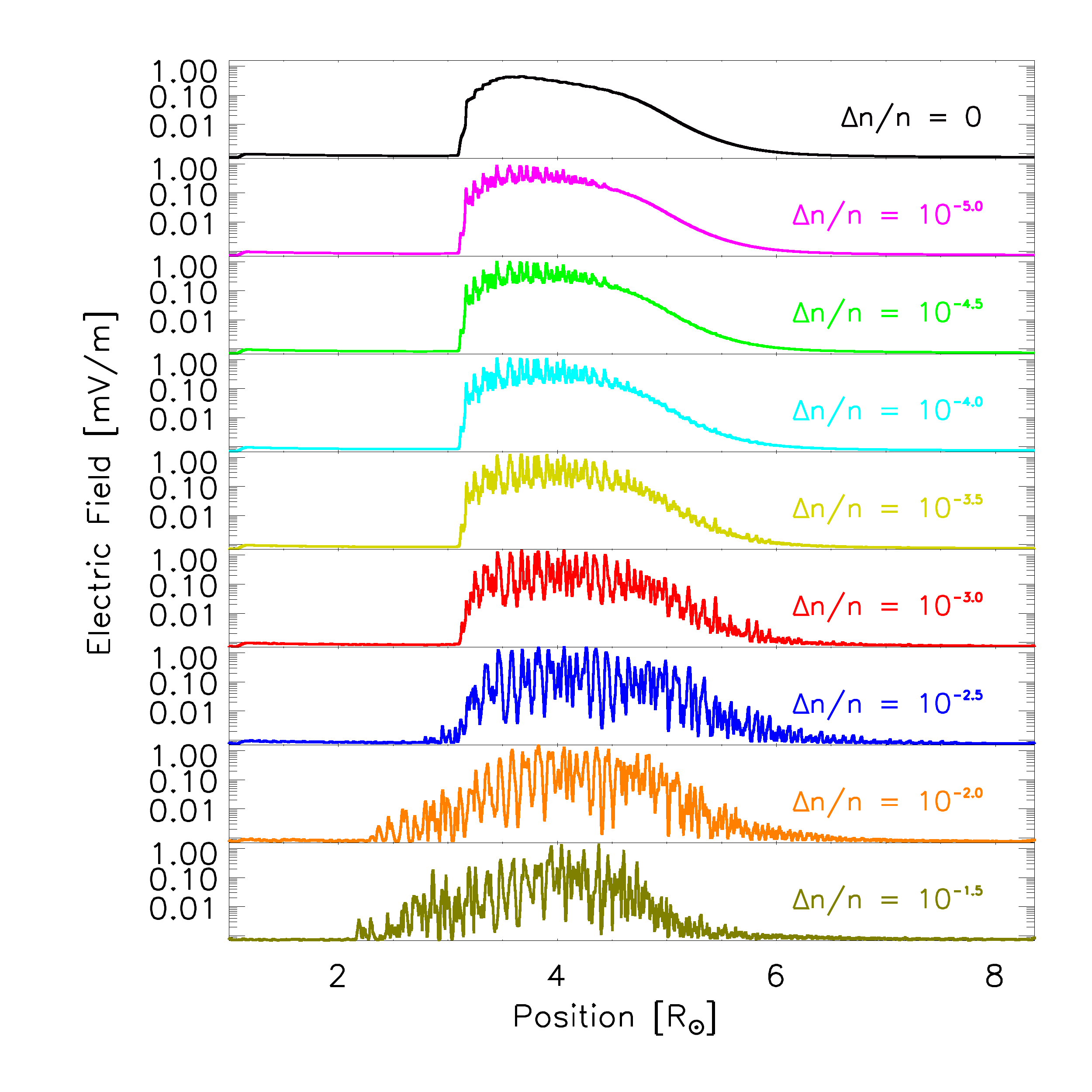}
  \includegraphics[width=0.99\columnwidth,trim=30 30 20 20,clip]{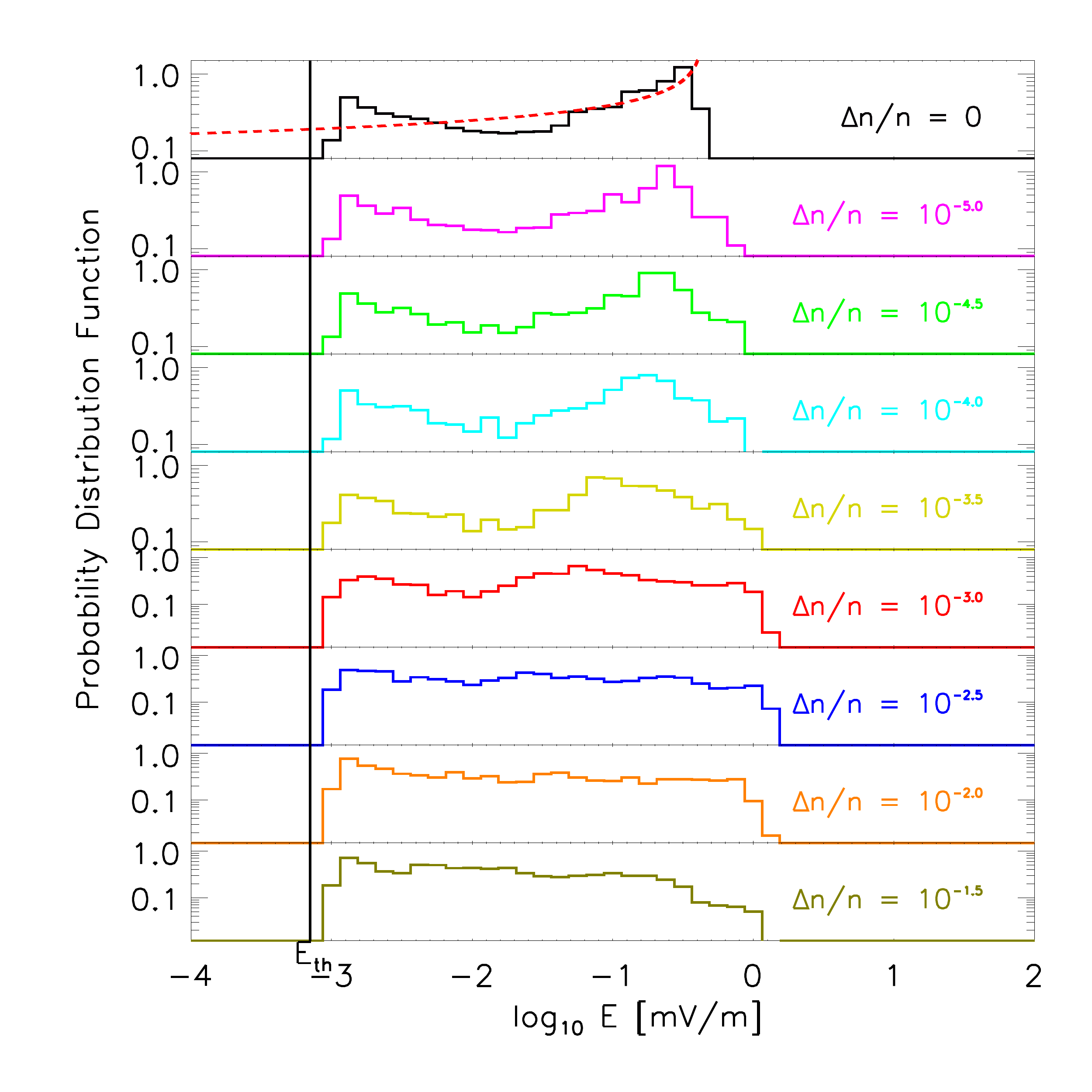}
\caption{Left, a: induced electric field from Langmuir waves produced from an electron beam propagating through a constant mean background density for different levels of density fluctuations after $t=277$~seconds of propagation.  The top panel has $\Delta n/n=0$.  The remaining panels increase $\Delta n/n$ from $\Delta n/n=10^{-5}$ to $\Delta n/n=10^{-1.5}$ in the bottom panel, as indicated on the right hand side.  Right, b: probability distribution functions of the electric field on the left. The bin size is 0.125 mV/m in log space.  The top panel has $\Delta n/n=0$ and the red dashed line represents the PDF of a Gaussian distribution.  The remaining panels increase $\Delta n/n$ from $\Delta n/n=10^{-5}$ to $\Delta n/n=10^{-1.5}$.  Note the change in y-axis limits at $\Delta n/n \geq 10^{-3}$.  The black vertical line indicates the thermal level of the electric field from Langmuir waves in plasma with $n_e=5~\rm{cm}^{-3}$.}
\label{fig:LW_Pos_Time_Earth}
\end{figure*}

The fluctuating component of the background plasma, described by Equation \ref{eqn:fluc}, is varied through the intensity of the density turbulence $\Delta n/n$.  Nine simulations were ran with $\Delta n/n$ from $10^{-1.5}, 10^{-2}, 10^{-2.5}, 10^{-3}, 10^{-3.5}, 10^{-4}, 10^{-4.5}, 10^{-5}$ and no fluctuations.  Propagation of the beam causes Langmuir waves to be induced after 80 seconds, relating to our choice of $\tau=20$~s.  Langmuir wave production increases as a function of time till around 200 seconds after which is remains roughly constant.

Figure \ref{fig:LW_Pos_Time_Earth}a shows a snapshot of the electric field from the Langmuir wave energy density after 277 seconds.  When $\Delta n/n=0$ (no fluctuations), the electric field has a smooth profile.  The wave energy density is dependent upon the electron beam density \citep{Melnik:2000aa} and so is concentrated in the same region of space as the bulk of the electron beam.  This region increase as a function of time as the range of velocities within the electron beam causes it to spread in space.  The electric field is smaller at the front of the electron beam where the number density of electrons is smaller.

When $\Delta n/n$ is increased the electric field shows the clumpy behaviour seen from in-situ observations.  At $\Delta n/n=10^{-2.5}$ and higher, the electric field is above the thermal level behind the electron beam.  The density fluctuations have refracted the Langmuir waves out of  phase speed range where they can interact with the electron beam.  Consequently these Langmuir waves cannot be re-absorbed as the back of the electron beam cloud passes them in space \citep{Kontar:2001ab}.  They are left behind, causing an energy loss to the propagating electron beam \citep[see][for analysis on beam energy loss]{Reid:2013aa}.

\subsection{Electric field distribution over the entire beam} \label{sec:beam_whole}

To analyse the distribution of the electric field over the entire beam we have plotted $P(\log E)$, the probability distribution function (PDF) of the base 10 logarithm of the electric field in Figure \ref{fig:LW_Pos_Time_Earth}b at $t=277$ seconds.  We have only considered areas of space where Langmuir waves were half an order of magnitude above the background level, or $\log[U_w/U_w(t=0)]>0.5$, to neglect the background from the PDF, corresponding to $E>1.78E_{th}$.  The PDF thus obeys the condition $\int_{1.78E_{th}}^{E_{max}}P(\log E)d\log E=1$.

The top panel in Figure \ref{fig:LW_Pos_Time_Earth}b shows $P(\log E)$ when $\Delta n/n=0$.  We have over-plotted the analytical PDF of a Gaussian (see Section \ref{sec:theory}) where $\sigma$ and $E_{max}$ were estimated from a fit to the simulation data.  The majority of the enhanced electric field is large in comparison to the thermal level and so $P(\log E)$ is focussed near the peak electric field around 0.3 mV/m, similar to the analytical PDF.  
$P(\log E) $ decreases as $E$ becomes smaller till around 0.01 mV/m.  Below 0.01 mV/m the enhanced electric field is produced by the front of the electron beam where the density of beam electrons is smaller.  The front of the beam is spread over a large region in space and consequently $P(\log E)$ increases for smaller values of $E$ till it reaches the thermal level.
Low electric fields at the front of the beam is consistent with the lack of observed Langmuir waves when electrons above 20~keV arrive at the spacecraft \citep{Lin:1981aa}.  $P(\log E)$ is effectively a combination of two components: the distribution at high electric fields from the bulk of the electron beam and the distribution at the low electric fields from the front of the beam.

The remaining panels in Figure \ref{fig:LW_Pos_Time_Earth}b show $P(\log E)$ when $\Delta n/n>0$.  As $\Delta n/n$ increases in value, $P(\log E)$ becomes less peaked at the highest values of $\log E$ and spreads out over a larger range in $\log E$.  The increase in $P(\log E)$ at small values of $\log E$ is present for all the simulations.  The increase in $\Delta n/n$ does not significantly alter the shape of $P(\log E)$ below 0.01 mV/m until the spreading of the electric field from $\Delta n/n$ becomes large.

\begin{figure}\center
	  \includegraphics[width=0.99\columnwidth]{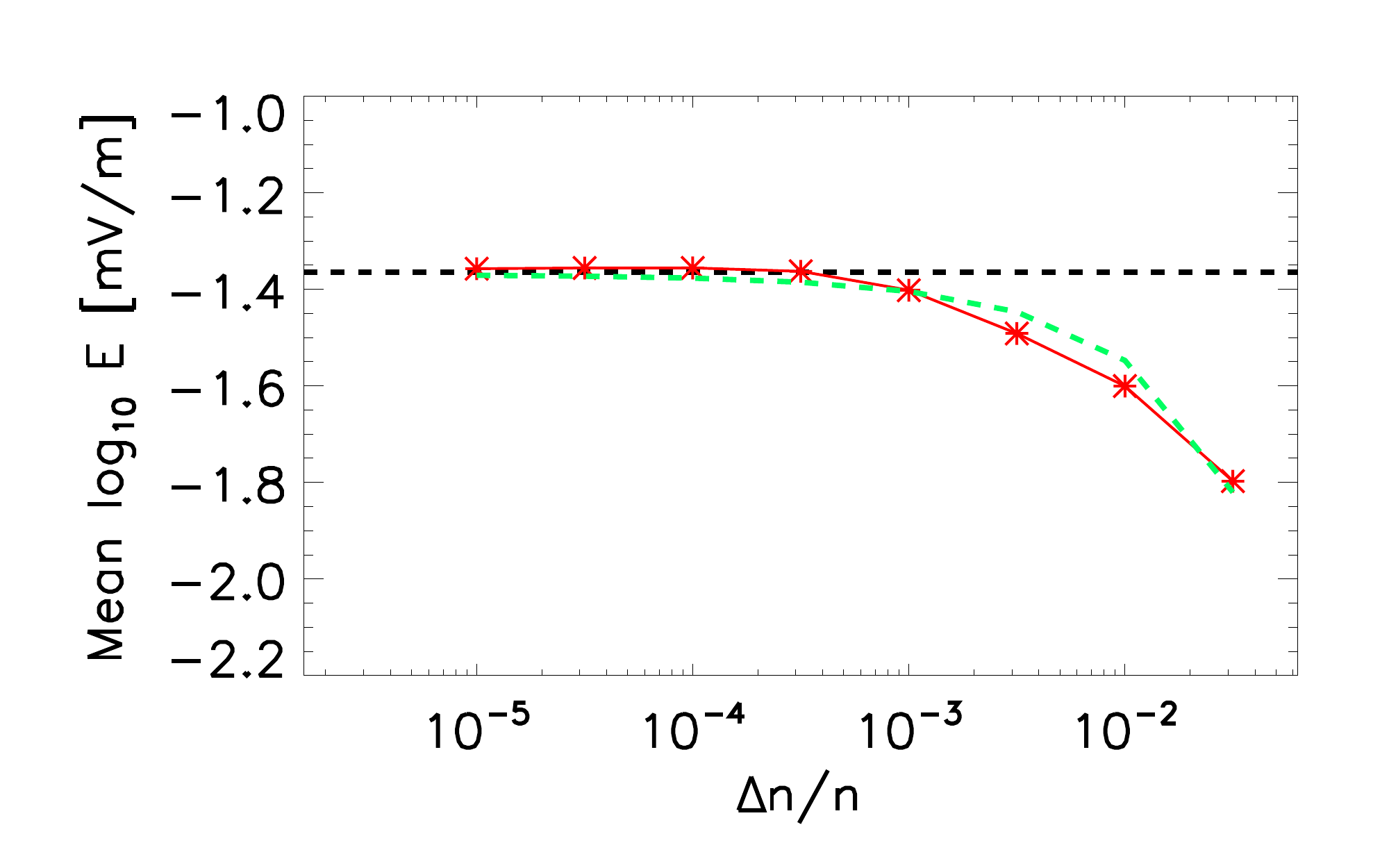}
  \includegraphics[width=0.99\columnwidth]{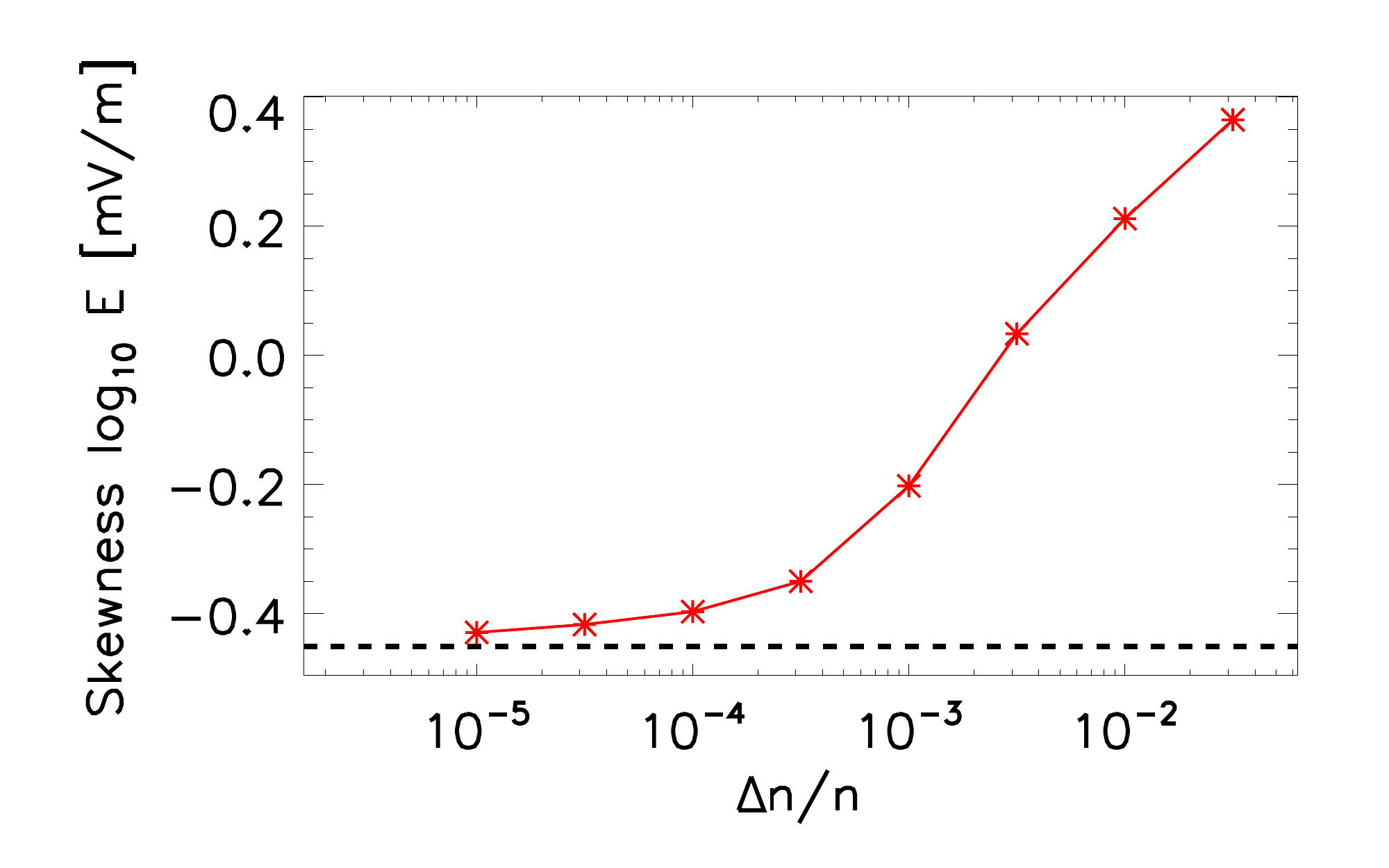}
\caption{The mean and skewness of the PDF of $\log_{10} E$ averaged between 200 to 310 seconds, plotted against $\Delta n/n$.  The horizontal black dashed line indicates the mean and skewness when $\Delta n/n=0$. A fit to the mean of $\log E$ using Equation \ref{eqn:fit_mean} is shown with a dashed green line.  For a discussion of the fit, see Section \ref{sec:discussion}.}
\label{fig:LW_Moments_Earth}
\end{figure}

To illustrate how the distribution of the electric field changes across the entire beam we have plotted the first and the third moment of $\log E$ as a function of $\Delta n/n$.    There are fluctuations but little systematic change in the moments of the electric field between 200--300 seconds so we averaged over this time range.  The mean of $\log E$ (first moment) characterises the average value whilst the skewness of $\log E$ (third moment) characterises the asymmetry of the distributions.  Both moments are plotted in Figure \ref{fig:LW_Moments_Earth} for all eight simulations when $\Delta n/n>0$, with the unperturbed ($\Delta n/n=0$) simulation represented as a horizontal black dashed line.  The standard deviation of $\log E$ (second moment) is not shown as there was little variation over the entire of the beam as a function of $\Delta n/n$.

The change in the mean of $\log E$ as a function of $\Delta n/n$ illustrates a decrease in the beam-induced electric field as $\Delta n/n$ increases.  The mean remains constant for weak density turbulence, despite the change in the shape of the distributions.  It is not until $\Delta n/n=10^{-3}$ that we see the mean of $\log E$ decrease significantly from the mean obtained in unperturbed plasma.  For $\Delta n/n=10^{-1.5}$ the density turbulence suppresses the mean of $\log E$ over the entire length of the beam by almost half an order of magnitude.

The change in skewness as a function of $\Delta n/n$ illustrates a shift in the electric field from being concentrated at the highest electric fields to the lowest electric field.  When $\Delta n/n=0$ the magnitude of the skewness of $\log E$ is high because the electric field distribution is more concentrated close to $E_{max}$; similar to the analytical PDF of a Gaussian presented in Section \ref{sec:theory}.  As $\Delta n/n$ increases, the skewness of $\log E$ decreases in magnitude and then changes sign.  When $\Delta n/n=10^{-1.5}$, $P(\log E)$ has only one peak near the thermal level, with a long tail to higher values of the electric field.

\subsection{Electric field distribution above a threshold value}  \label{sec:beam_above_thresh}

As mentioned previously, there are two components to the probability distribution of $\log E$, one at high electric fields from the bulk of the electron beam and one at low electric fields from the front of the electron beam.  For the simulations where $\Delta n/n \leq 10^{-3}$ there is a noticeable change in the shape of $P(\log E)$ above 0.01 mV/m despite the mean of $\log E$ remaining relatively constant.  We analyse how the distribution above 0.01 mV/m varies as the highest electric fields are indicative of what is measured in-situ by spacecraft in the solar wind.

\begin{figure}\center
  \includegraphics[width=0.99\columnwidth]{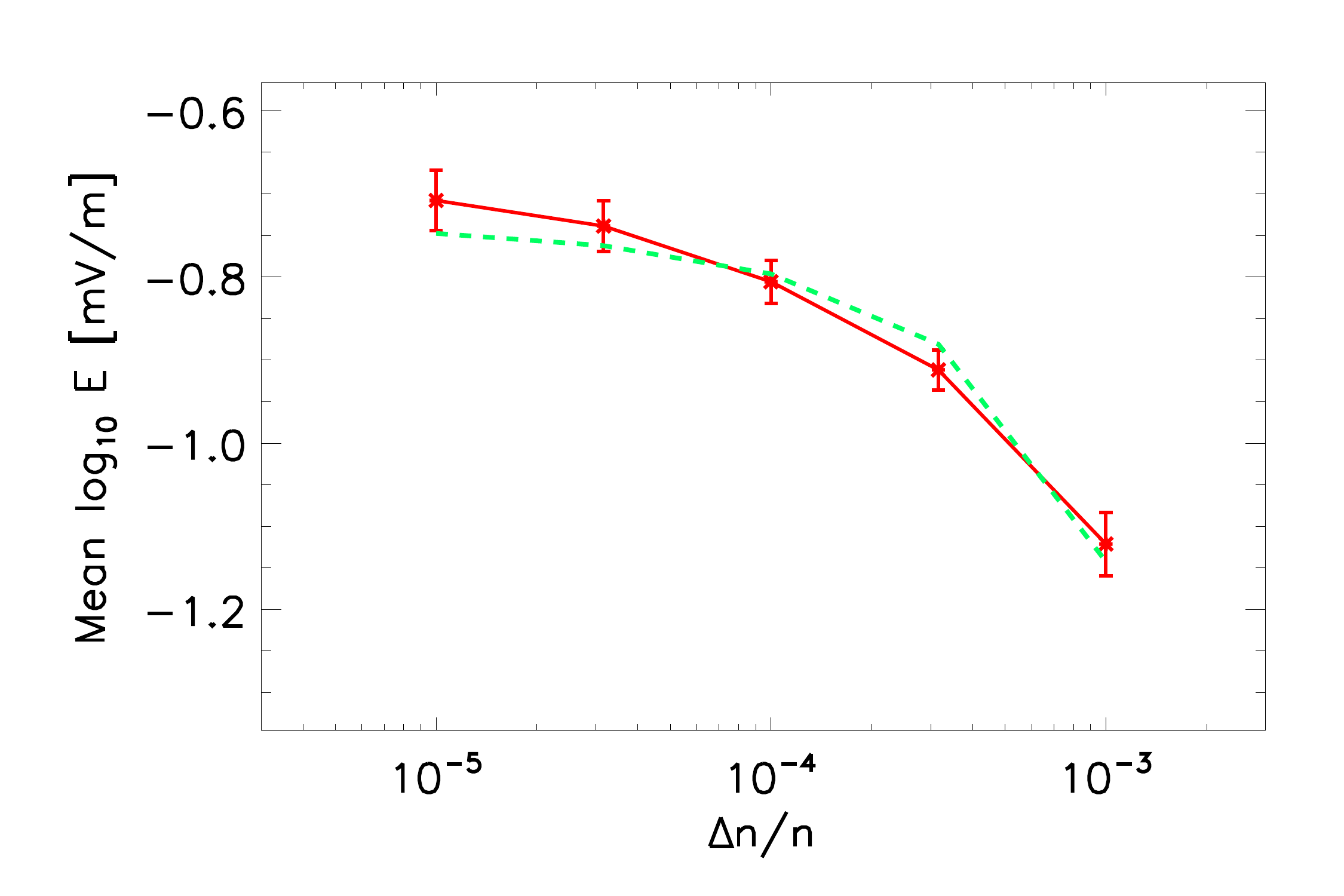}
  \includegraphics[width=0.99\columnwidth]{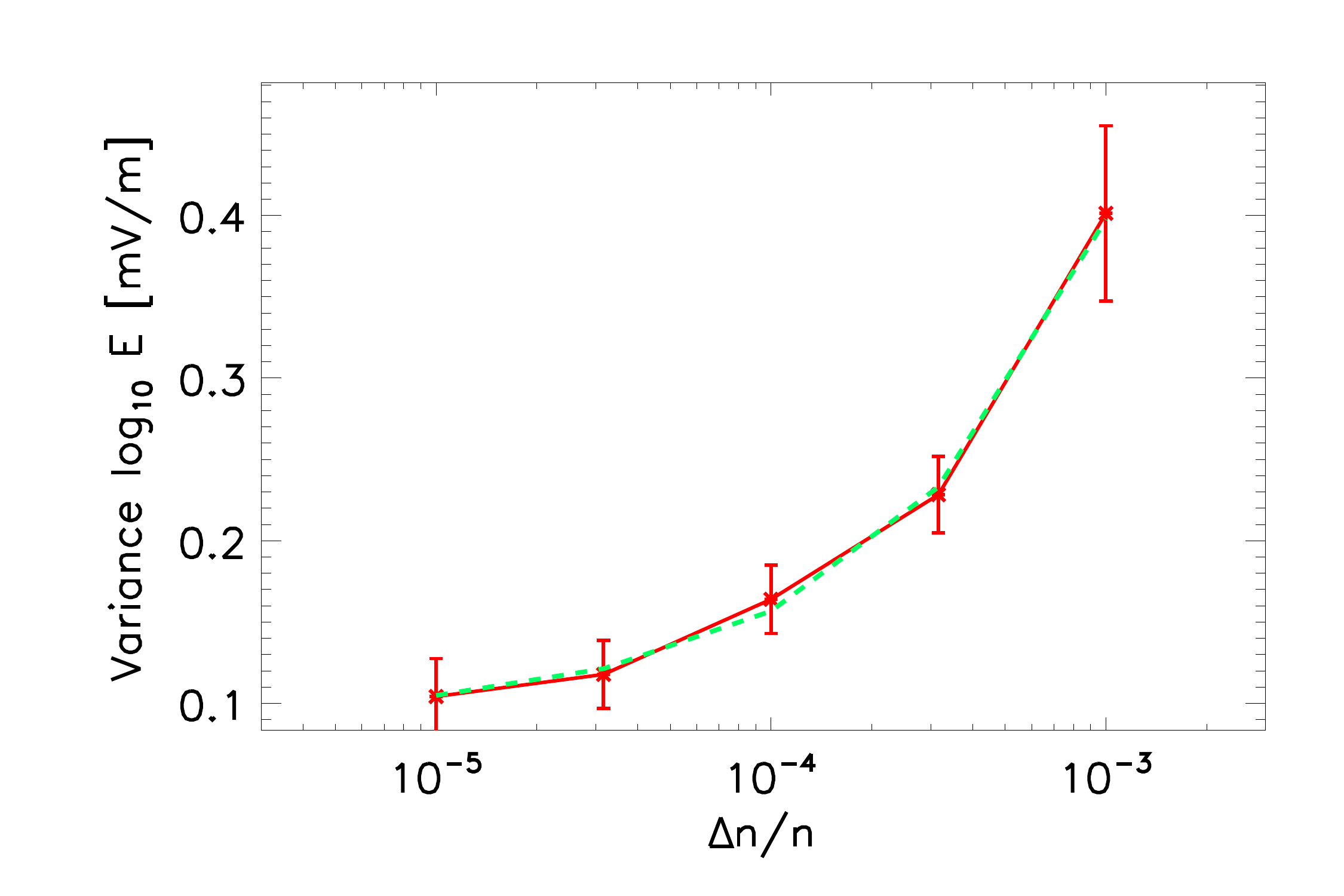}
\caption{The mean and variance from a Gaussian fit to $P(\log~E)$ above 0.01 mV/m as a function of $\Delta n/n$.  A fit to the data is shown by the dashed green line using Equation \ref{eqn:fit_mean} for $c_3=2/3$.  The inverse of Equation \ref{eqn:fit_mean} is shown over the log-variance by the green dashed line for $c_3=2/3$.  For a discussion of the fits, see Section \ref{sec:discussion}.}
\label{fig:LW_Moments_High_Earth}
\end{figure}

To obtain an estimate of the most likely value and the spread in $\log E$ we modelled $P(\log E)$ above 0.01 mV/m with a Gaussian distribution to find the mean and variance.  The Gaussian distribution is only an approximation of the distribution, especially when $\Delta n/n=0$ (see Figure \ref{fig:LW_Pos_Time_Earth}b) but provides fit parameters that highlight a general trend.

We find the mean of the Gaussian fit decreases as $\Delta n/n$ increases, shown in Figure \ref{fig:LW_Moments_High_Earth}b.  This can be seen visually in Figure \ref{fig:LW_Pos_Time_Earth}b by the mode of the distribution decreasing as $\Delta n/n$ increases.  As the distribution of the electric field becomes more clumped in space, the mean value (in log space) of each clump decreases.  Conversely, the variance of the Gaussian fit increases as $\Delta n/n$ increases.  Again this can be seen visually in Figure \ref{fig:LW_Pos_Time_Earth}b by the larger spread in the electric field.  The increase in the variance mirrors the decrease in the mean and occurs at a similar rate.

\subsection{Electric field distribution over a subset of the beam} \label{sec:beam_peak}

We further highlight how increasing $\Delta n/n$ causes a decrease in the mean and an increase in the variance of $\log E$ by plotting a subset of the beam.  The subset used is a length of the beam where the most intense Langmuir waves are observed when $\Delta n/n=0$.  We use the condition $U_w/U_w(t=0) \geq 10^5$, corresponding to electric fields above 0.22 mV/m and a region of space at $t=277$ s that is 0.9 solar radii in length.  The length increases as a function of time as velocity dispersion stretches the electron beam.  Sampling this subset of the beam removes the contribution to the electric field from the front of the beam.

\begin{figure}\center
  \includegraphics[width=0.99\columnwidth,trim=50 30 20 20,clip]{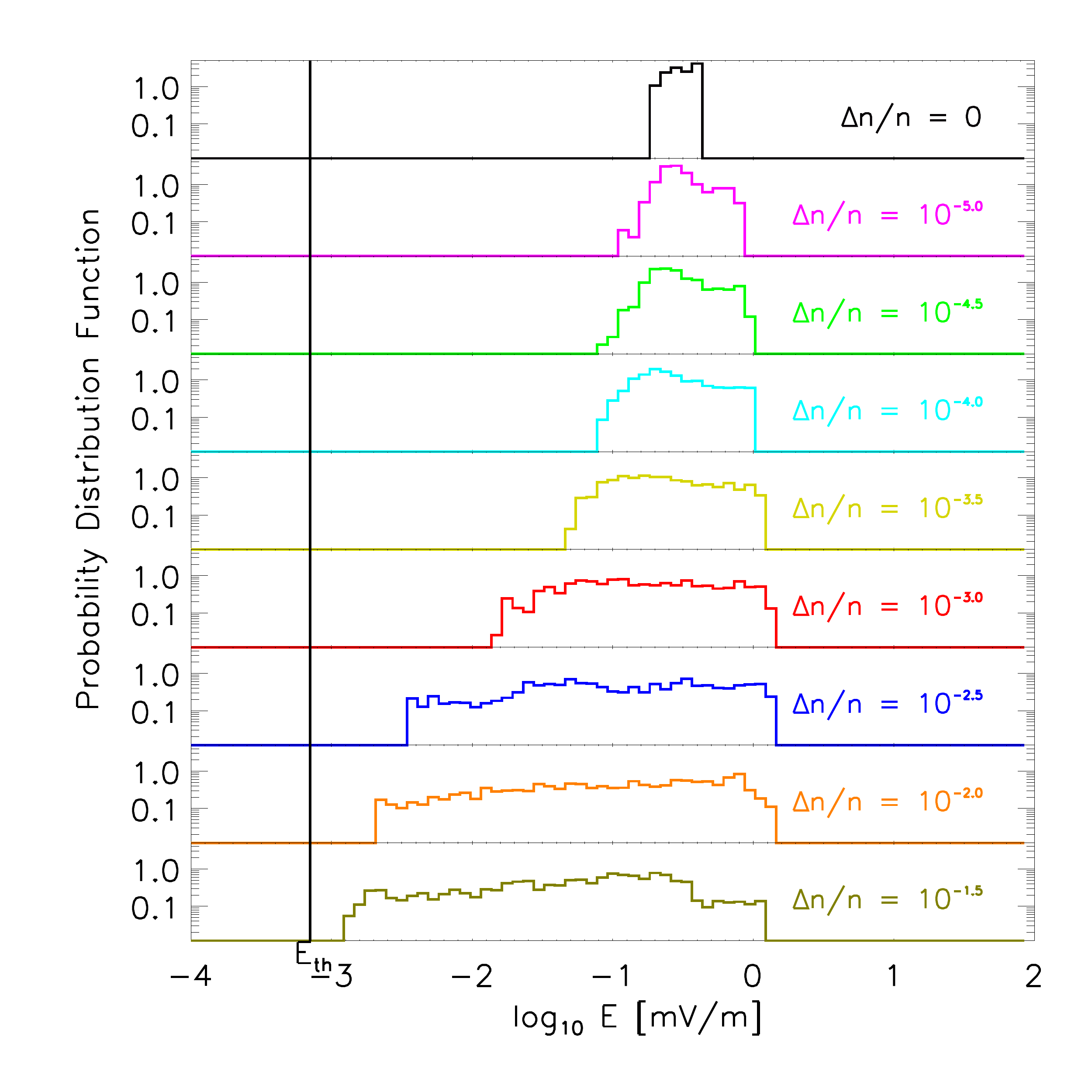}
\caption{Probability distribution function of the electric field at $t=277$~seconds that is induced from the central part of the electron beam where $E>0.22$~mV/m in the unperturbed case.  The top panel has $\Delta n/n=0$.  The remaining panels increase $\Delta n/n$ from $\Delta n/n=10^{-5}$ to $\Delta n/n=10^{-1.5}$ in the bottom panel, as indicated on the right hand side.  The black vertical line indicates the thermal level of the electric field from Langmuir waves.}
\label{fig:LW_SGT_Earth_PDF}
\end{figure}

Figure \ref{fig:LW_SGT_Earth_PDF} shows $P(\log E)$ over this subset of the electron beam for the different values of $\Delta n/n$ at $t=277$~s.  When $\Delta n/n=0$ the distribution is narrow and only above 0.22 mV/m, as defined from the sampling condition.  As the level of fluctuations increases the clumping in the electric field causes the distribution to be spread over a larger range of values.  The width of the distribution increases till it gets close to the thermal level for the highest values of $\Delta n/n$.  The form of the distribution is noticeably different to $P(\log E)$ sampled over the entire beam and highlights the change to the electric field distribution from the presence of density fluctuations. The mean and the variance of $\log E$ behave in a similar manner to Figure \ref{fig:LW_Moments_High_Earth} when $\Delta n/n <10^{-2}$.  When $\Delta n/n=10^{-2},10^{-1.5}$ the spreading of $\log E$ reaches the thermal level and so the variance cannot continue to increase at the same rate.  The mean subsequently decreases at a slower rate.

\section{Electron beams at the Sun} \label{sec:Sun}

We now explore the evolution of the electric field from a propagating electron beam that is injected into the low corona and propagates out of the solar atmosphere.  A key difference to the previous section is that electron beams travelling through the coronal and solar wind plasma experience the large-scale decrease in the background electron density.  The parameter regime is also different for an electron beam at the Sun, with higher density beams, higher energy electrons in the beam interacting resonantly with Langmuir waves and higher background electron densities.

We model the large-scale decrease in the background density $n_0(r)$ and the small-scale density fluctuations as described in Section \ref{sec:plasma} except that we also increase the level of density turbulence as a function of distance from the Sun.  A smaller value of $\Delta n/n$ closer to the Sun is observed by scintillation techniques \citep{Woo:1995aa,Woo:1996aa} and Helios in-situ measurements of the fast solar wind \citep{Marsch:1990aa}.  We model $\Delta n/n$ changing with distance according to
\begin{equation}\label{eqn:fluc_rad}
\frac{\Delta n}{n}(r) = \sqrt{\frac{\langle \Delta n(r)^2 \rangle}{\langle n(r) \rangle^2}} = \left(\frac{n_0(1 AU)}{n_0(r)}\right)^{\Psi} \sqrt{\frac{\langle \Delta n(r=1~AU)^2 \rangle}{\langle n(r=1~AU) \rangle^2}}
\end{equation}
where $\Psi=0.25$, derived from the results of \citet{Reid:2010aa} based upon the ratio of the electron spectral index above and below the break energy observed in simulations after reaching 1~AU ($214~R_\odot$).  This results in a level of fluctuations at the Sun that is roughly $1\%$ of the level at the Earth, or $\frac{\Delta n}{n}(\rm{Sun})=10^{-2}\frac{\Delta n}{n}(1~\rm{AU})$.

\vspace{20pt}
\begin{center}
\begin{table*}
\centering
\begin{tabular}{ c  c  c  c  c  }

\hline\hline

Energy Limits & Velocity Limits & Spectral Index & Temporal Profile & Density Ratio \\ \hline
1.4~eV to 113~keV &  $4-36~v_{th}$ & $\alpha=8.0$& $\tau=1$~s & $n_b/n_e=10^{-5}$ \\

\hline
\end{tabular}
\vspace{20pt}
\caption{Initial beam parameters for the electron beam injected into the solar corona.}
\label{tab:beam_sun}
\end{table*}
\end{center}

We set the injection height in the corona to $3\times10^9$~cm [$0.04~R_\odot$], corresponding to a background density of $n_{e}=10^{9.5}~\rm{cm}^{-3}$.  The background plasma temperature was set to $2\times10^6$~K, indicative of the flaring solar corona.  The beam parameters are given in Table \ref{tab:beam_sun}.  The velocity limits are within the range of exciter velocities derived from the drift rate of type III bursts.  The spectral index is typical of electron spectra inferred from hard X-ray measurements \citep{Holman:2011aa}.  The size of the electron beam $d=10^9$~cm [$0.014~R_\odot$], inferred as a typical size of a flare acceleration region \citep{Reid:2014aa}.  The time injection is indicative of a type III duration at high frequencies around 400~MHz.  The ratio of beam and background density is large enough that a substantial number of electrons are injected into the system but small enough that the beam will not alter the background Maxwellian distribution function.

\subsection{Beam-induced electric field}

\begin{figure*}\center
  \includegraphics[width=0.99\columnwidth,trim=20 30 0 20,clip]{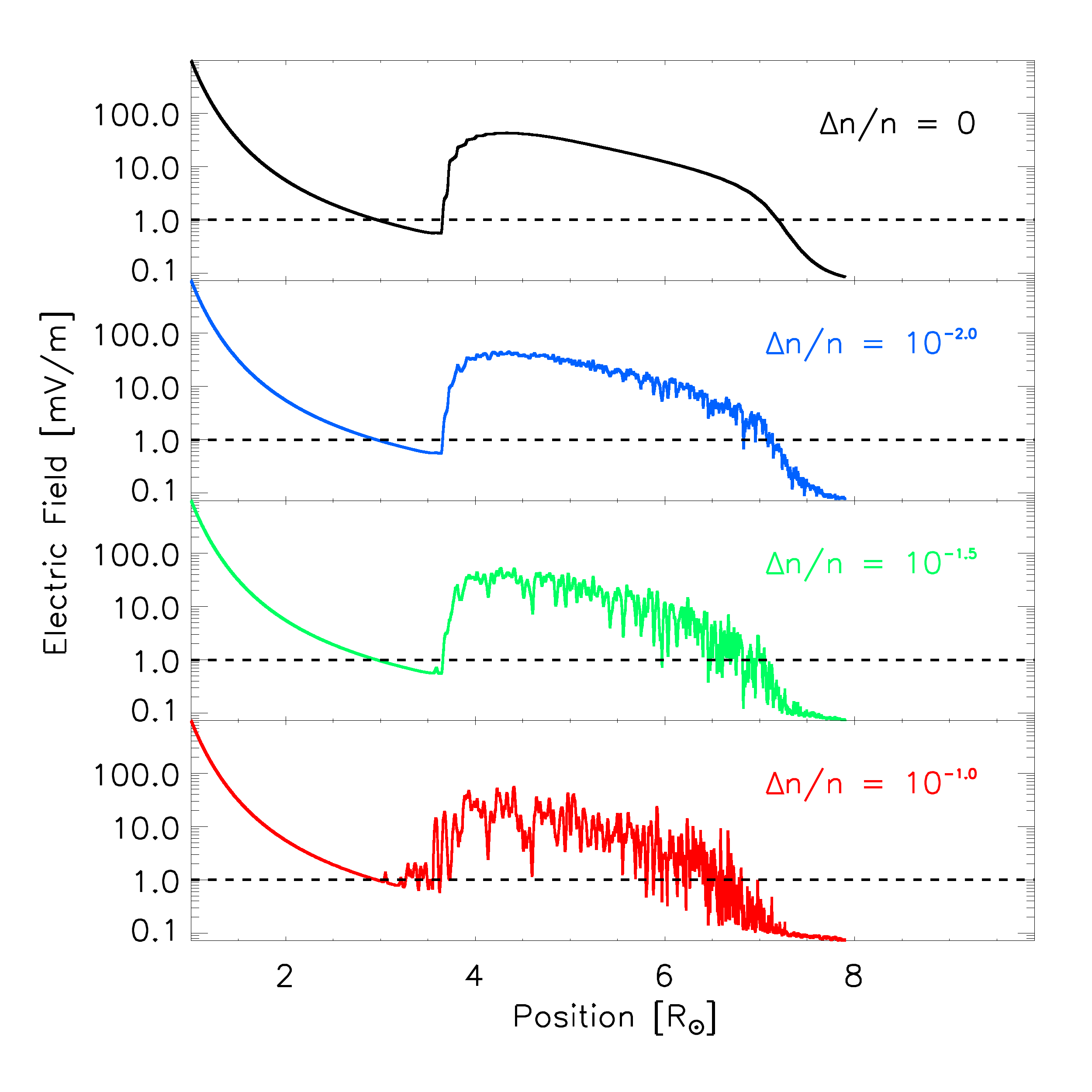}
  \includegraphics[width=0.99\columnwidth,trim=20 30 0 20,clip]{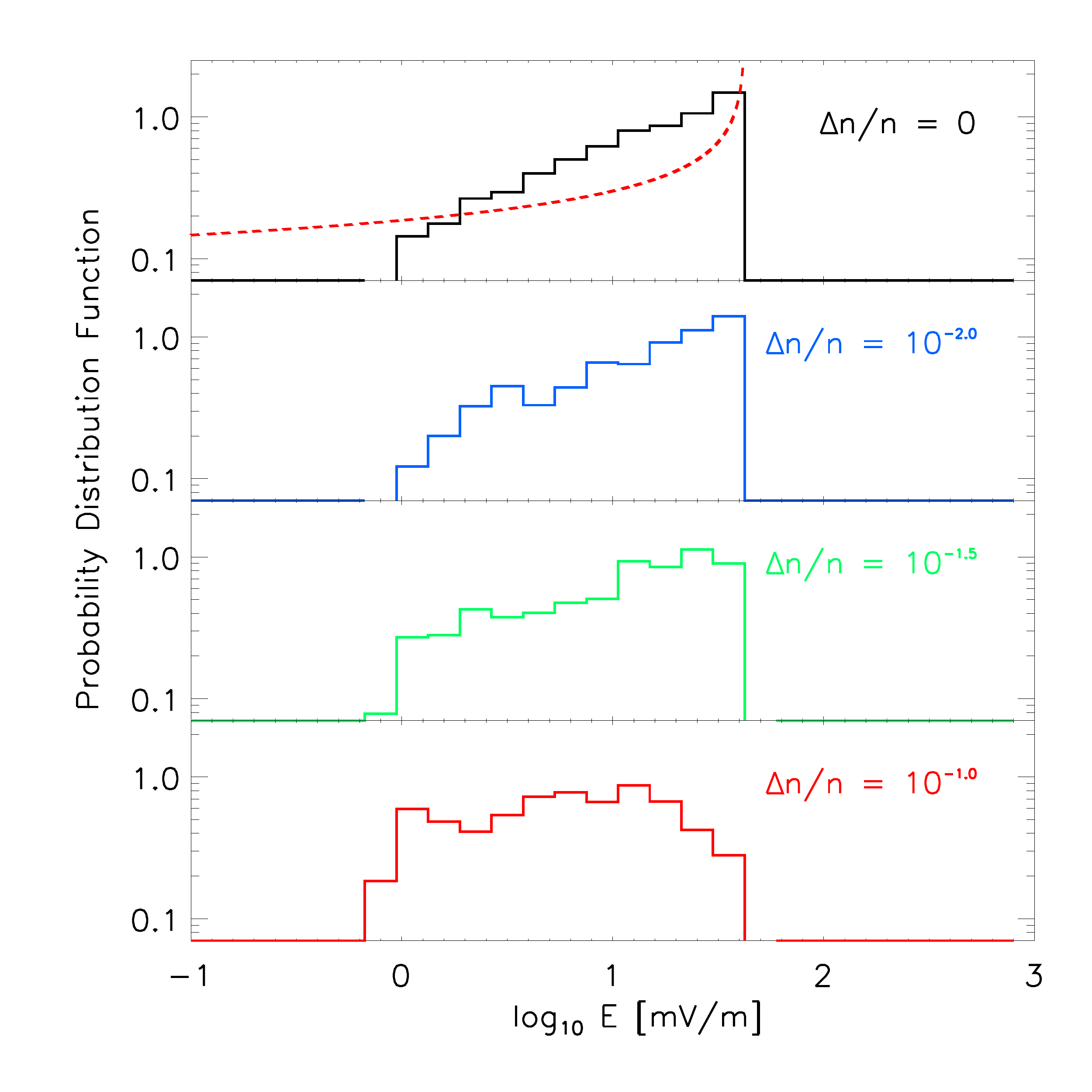}
\caption{Left, a: electric field induced from an electron beam injected in the solar corona and propagating out from the Sun for different levels of $\frac{\Delta n}{n}(r)$ with a decreasing mean background electron density, normalised at $\frac{\Delta n}{n}(1~\rm{AU})$.  The top panel has $\frac{\Delta n}{n}(r)=0$.  The remaining panels are normalised by $\frac{\Delta n}{n}(1~\rm{AU})=10^{-2}, 10^{-1.5}, 10^{-1}$ from top to bottom.  The black dashed line indicates $E=1$~mV/m used as a lower limit for the probability distribution function.  Right, b: probability distribution functions of $\log E$. The bin size is 0.15 in log space.  The top panel has $\Delta n/n(r)=0$ where the red dashed line representing the PDF of a Gaussian fit with similar mean and standard deviation.  The remaining panels have $\frac{\Delta n}{n}(r)$ normalised by $\Delta n/n~[1~\rm{AU}]=10^{-2}, 10^{-1.5}, 10^{-1}$ from top to bottom.}
\label{fig:LW_Hist_Sun}
\end{figure*}

The magnitude of $\frac{\Delta n}{n}(r)$ decreases as $r\rightarrow0$ close to the Sun (Equation \ref{eqn:fluc_rad}).  $\frac{\Delta n}{n}(r)$ is normalised by the value chosen at $\frac{\Delta n}{n}(1~\rm{AU})$.  To explore how the intensity of density fluctuations influences the distribution of the electric field we varied $\frac{\Delta n}{n}(r)$ such that $\frac{\Delta n}{n}(1~\rm{AU})=10^{-1},10^{-1.5},10^{-2}$ and no fluctuations.  In all simulations, Langmuir waves are induced after 4 seconds, related to our choice of $\tau=1$~s.

The electric field induced from Langmuir wave growth as a function of position is shown in Figure \ref{fig:LW_Hist_Sun}a after 50 seconds of propagation.  The background electric field decreases with distance, corresponding to the decrease of the mean background electron density.  The increased level of inhomogeneity causes the Langmuir waves to be excited in clumps.  When we set $\frac{\Delta n}{n}(1~\rm{AU}) = 10^{-1}$ the tail of the electron beam is no longer able to fully reabsorb all the excited Langmuir waves due to increased wave refraction, in a similar manner to what occurred in Section \ref{sec:Earth}.  The electric field at 3.5 $R_{\odot}$ from the Sun is noticeably above the background compared to the simulation with zero fluctuations, where the tail of the electron beam has reabsorbed all the induced Langmuir waves.

\subsection{Electric field distribution}

As the background level of the electric field varies as a function of distance we consider the probability distribution of the electric field above 1~mV/m at $r>3~R_{\odot}$ such that $\int_{1}^{E_{max}}P(\log E)d\log E=1$.  We display $P(\log E)$ in Figure \ref{fig:LW_Hist_Sun}b after 50 seconds of propagation.  When $\Delta n/n=0$ the distribution is peaked at the highest field values.  The red dashed line is the analytical PDF of a Gaussian, described by Equation \ref{eqn:gauss_pdf} with $\sigma$ and $E_{max}$ approximated by a fit to the data.  Whilst the distribution of the electric field agrees with the analytical PDF insofar as it is peaked at the highest values, the analytical PDF fails to capture the rate of the decrease in the distribution.  This is because a decreasing background level of the electric field is not accounted for in Equation \ref{eqn:gauss_pdf}.

For the simulations where $\Delta n/n > 0$ the distributions become less peaked at the highest electric fields and more evenly distributed over $\log E$, in a similar way as was demonstrated in Section \ref{sec:Earth}.  Sampling the electric field above 1~mV/m means that we do not see the low intensity component of the electric field from the front of the electron beam.

\begin{figure}\center
  \includegraphics[width=0.99\columnwidth]{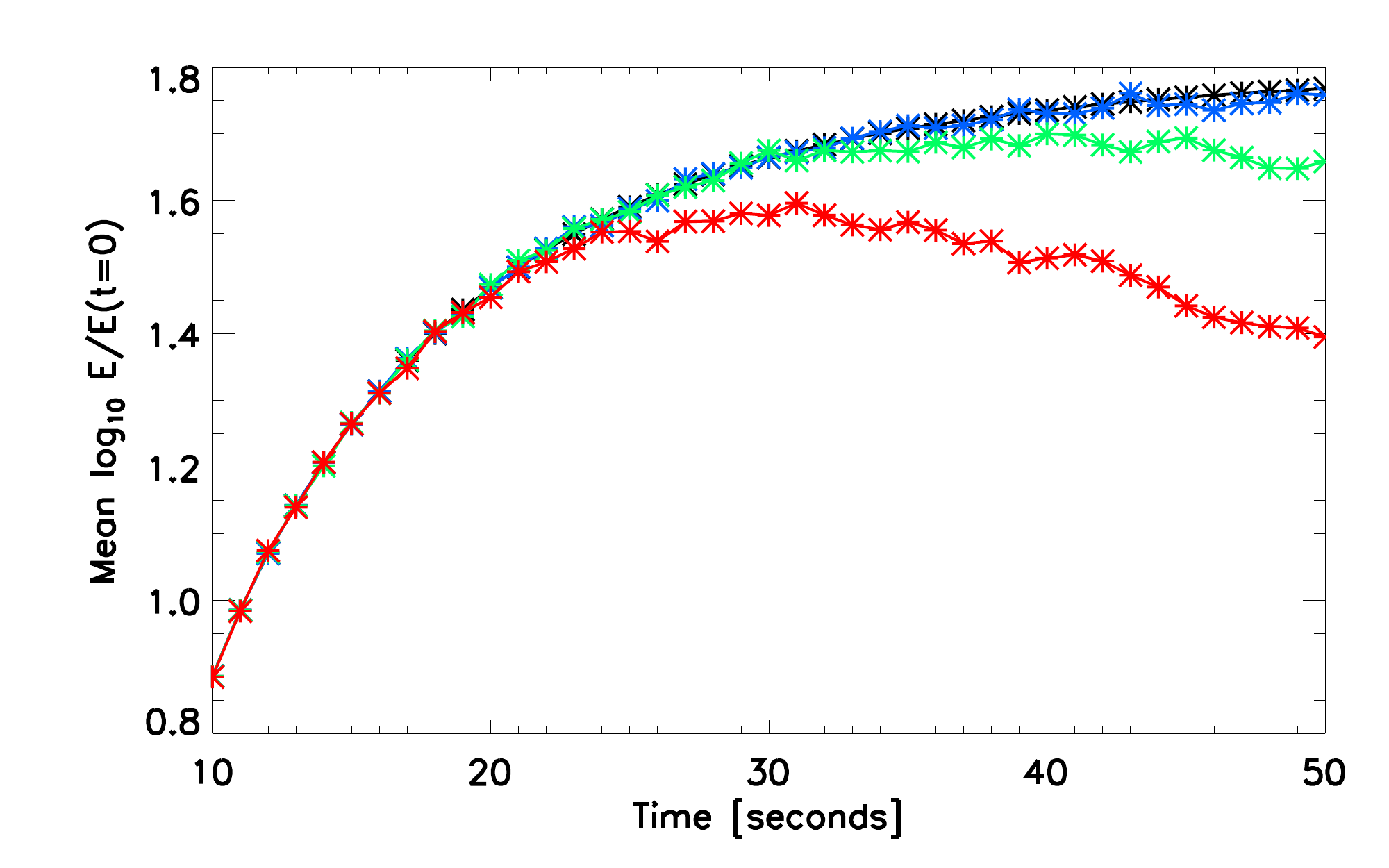}
  \includegraphics[width=0.99\columnwidth]{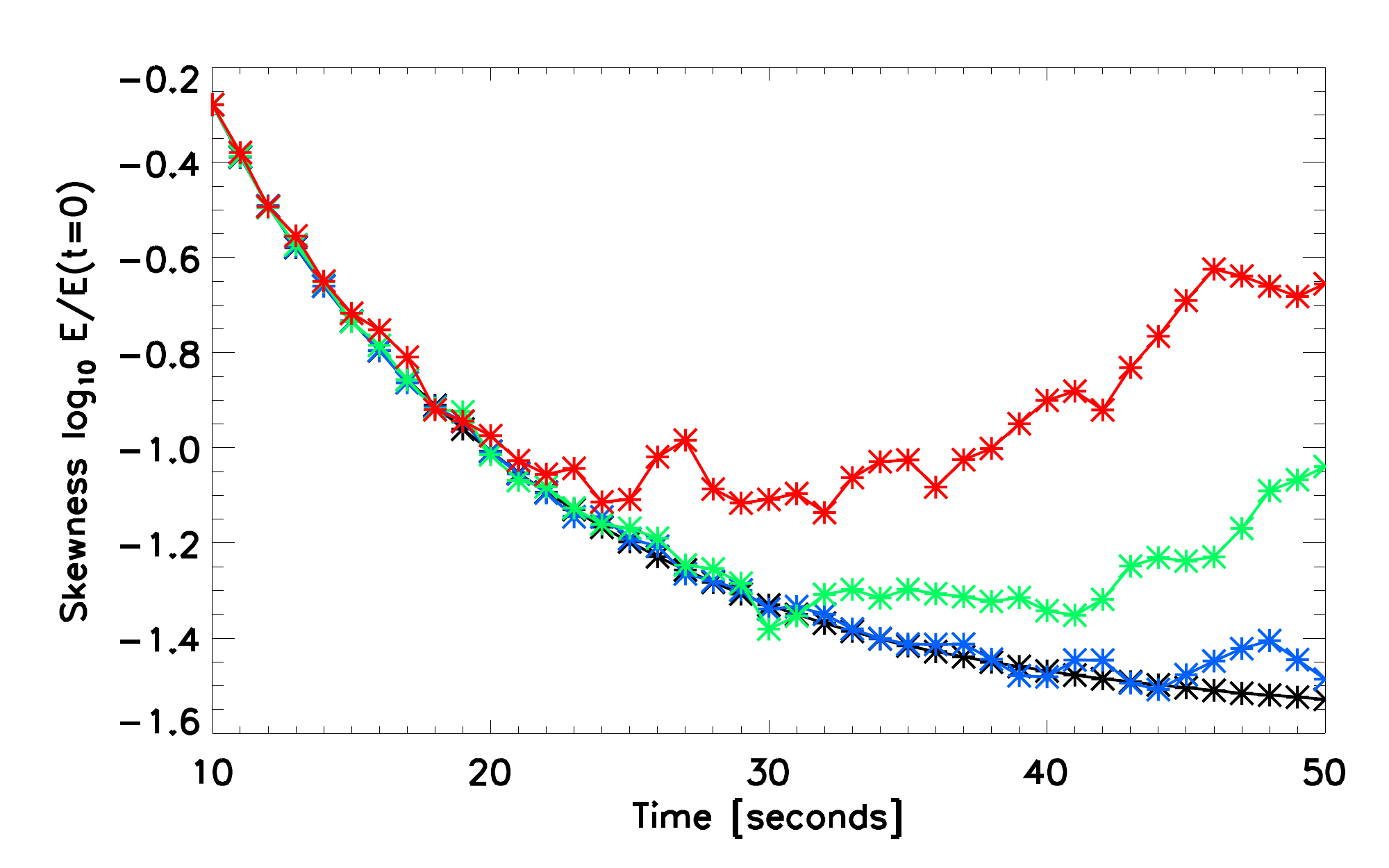}
\caption{Time dependence of the mean and skewness of the log normalised electric field $E/E(t=0)$ between 10 to 50 seconds.  Different colours relate to different levels of density fluctuations, similar to Figure \ref{fig:LW_Hist_Sun}.  The electron beam takes around 10 seconds before it start to produce significant levels of Langmuir waves.}
\label{fig:LW_Moments_Sun_Time}
\end{figure}

\subsection{Electric field moments}

For a beam travelling through the solar corona, the moments of $\log E$ vary as a function of time due to a number of effects including the decreasing background mean density, the radial expansion of the field and the changing $\Delta n/n(r)$.  The time dependence of the normalised mean of $\log E$ and the skewness of $\log E$ are shown in Figure \ref{fig:LW_Moments_Sun_Time}, characterising the average value and the asymmetry of the distribution respectively.  We show the normalised mean and skewness ($E/E(t=0)$) to remove the effect of the decrease in the background density as a function of distance from the Sun.  The decrease in background density does not significantly affect the asymmetry of the distribution but it dominates the behaviour of the mean electric field.

Close to the Sun the normalised mean of $\log E$ initially increases for all simulations as a function of time.  The increase continues for the duration of the simulation when $\frac{\Delta n}{n}(r)=0$.  For the higher values of $\frac{\Delta n}{n}(r)$ the normalised mean of $\log E$ stops increasing and begins to decrease at earlier times, corresponding to distances closer to the Sun.  The peak in the normalised mean of $\log E$ relates to the peak in Langmuir wave growth from the electron beam and occurs as early as 30 seconds when $\frac{\Delta n}{n}(1~\rm{AU})=10^{-1}$.  This result insinuates that the level of density fluctuations plays a significant role in determining when a solar electron beam produces the peak electric field above the thermal level and could play a significant role in determining which radio frequency of a type III burst has the highest flux.

The skewness of $\log E$ initially increases in magnitude as a function of time.  The increase in the asymmetry corresponds to an increase in the tail of the distribution at low E and is caused by the electron beam spreading out in space, producing electric fields with a lower magnitude.  In Figure \ref{fig:LW_Moments_Sun_Time} where $\frac{\Delta n}{n}(1~\rm{AU}) \geq 10^{-1.5}$ (red and green line), the skewness of $\log E$ begins to decrease in magnitude at the same time as the normalised mean of $\log E$ begins to decrease.  The change in skewness highlights the distribution becomes less concentrated at the highest electric fields and becoming more uniform, evident in Figure \ref{fig:LW_Hist_Sun}b when $\frac{\Delta n}{n}(1~\rm{AU})=10^{-1}$.

\section{Electron beams in the inner heliosphere} \label{sec:helio}

The current upcoming missions of Solar Orbiter and Solar Probe Plus will provide the opportunity to obtain in-situ measurements of the inner heliosphere that can test our theories about how the electric fields associated with propagating electron beams develop as a function of distance.  We therefore extended one simulation out to 0.34~AU or 75 solar radii.  Due to computational constraints we only ran one simulation out to this distance.  We normalised $\frac{\Delta n}{n}(r)$ using a value of $\frac{\Delta n}{n}(1~\rm{AU})=10^{-2}$, indicative of observed values \citep{Celnikier:1987aa}.  The initial parameters are the same as Section \ref{sec:Sun} except we increased the value of $n_{beam}=10^5~\rm{cm}^{-3}$ but also increased the temporal injection to $\tau=10$~seconds to represent a longer duration flare.

\subsection{Beam-induced electric field}

\begin{figure*}\center
  \includegraphics[width=0.99\columnwidth]{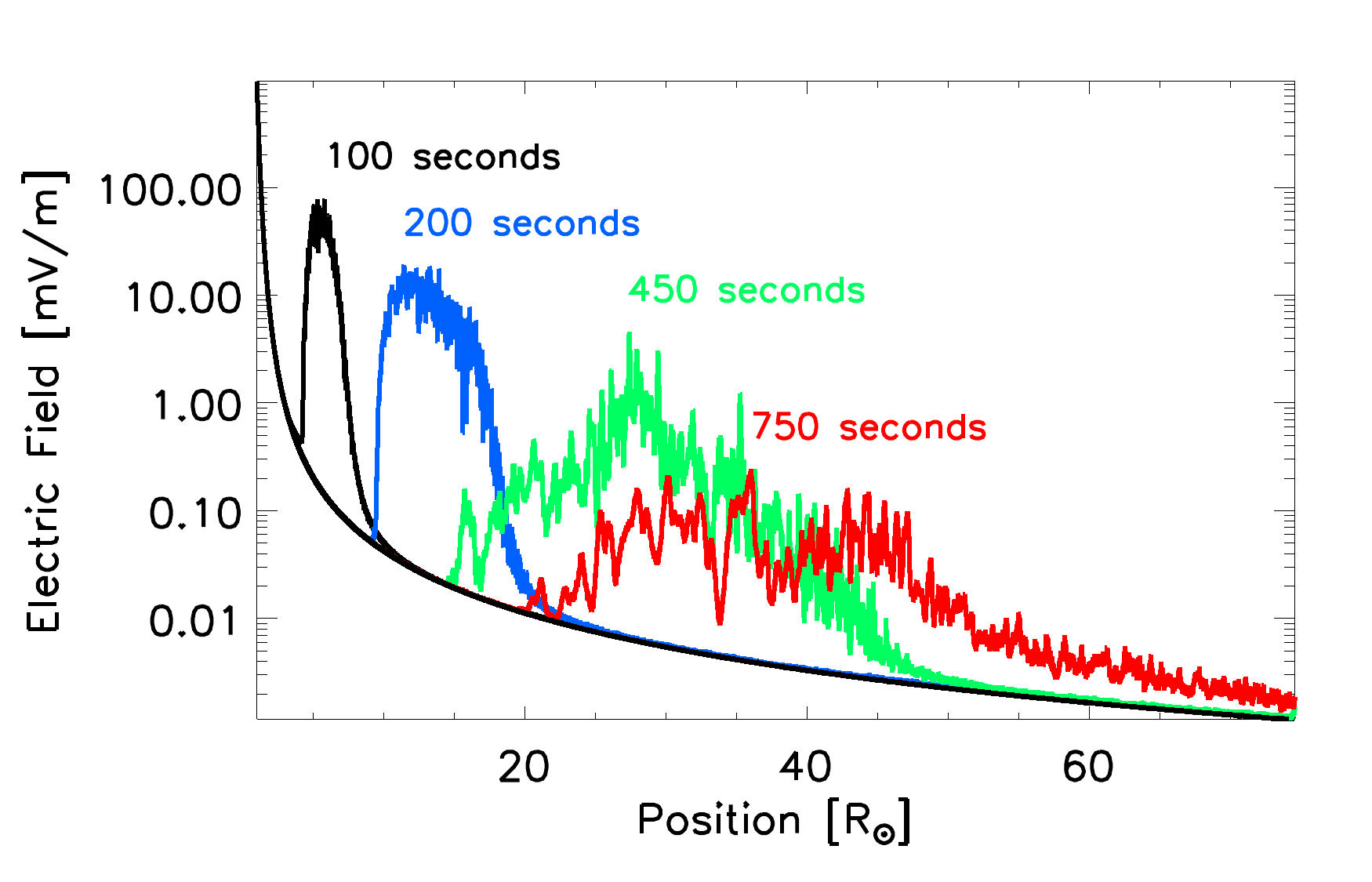}
  \includegraphics[width=0.99\columnwidth]{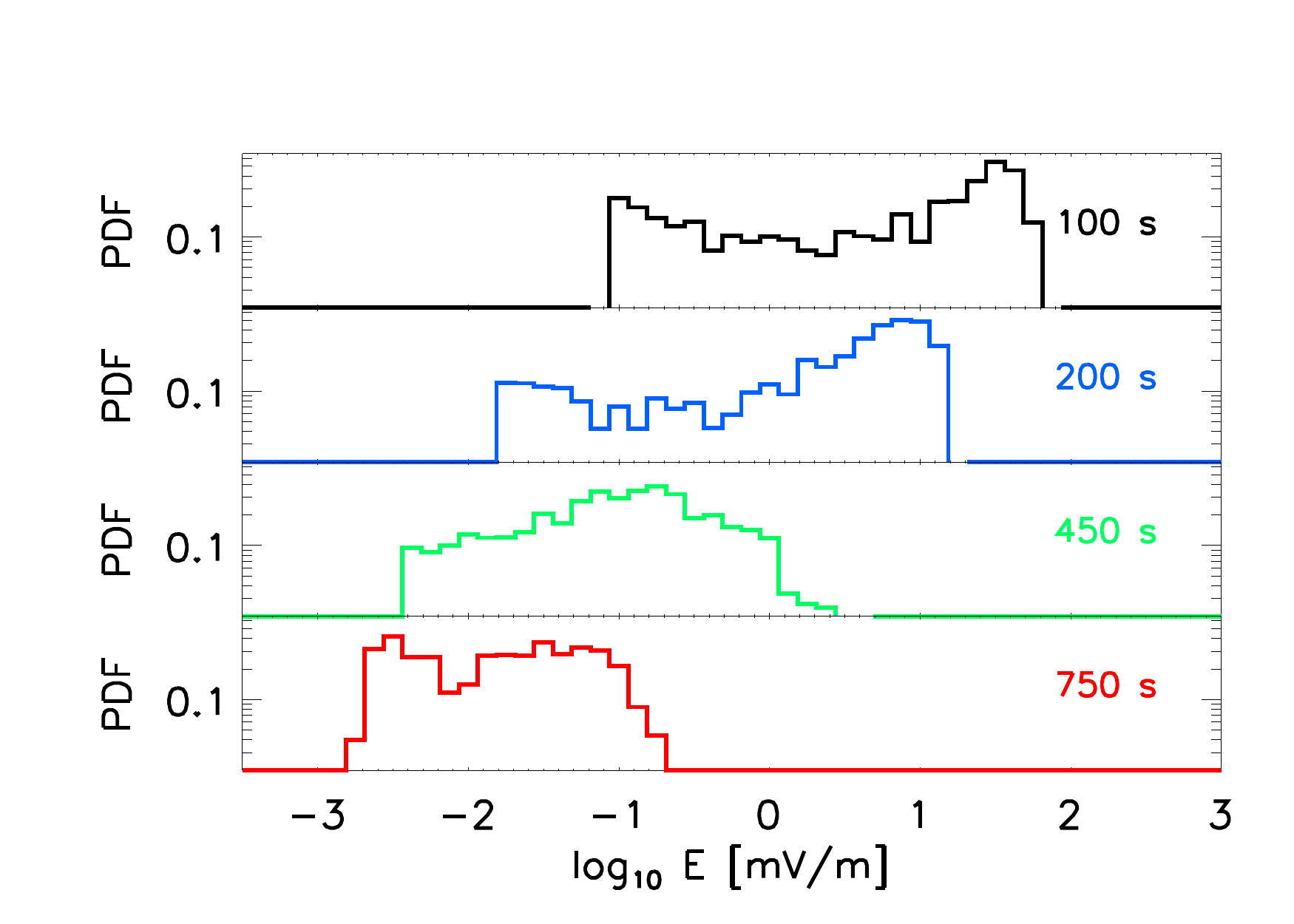}
\caption{Left a: electric field as a function of distance induced from an electron beam injected in the solar corona and propagating out through the solar corona and the solar wind.  The electric field is displayed at 100, 200, 450 and 750 seconds after injection.  Right b: probability density function of $\log E$ for 100, 200, 450, 750 seconds after electron injection from top to bottom panel.  The bin size is 0.125 in log space.}
\label{fig:LW_Pos_Time_Helio}
\end{figure*}

Snapshots of the electric field as a function of distance are shown in Figure \ref{fig:LW_Pos_Time_Helio}a at different times after beam injection.  At $t=100$~seconds the density turbulence is weaker closer to the Sun and the beam has a limited radial spread in space.  The resultant electric field is three orders of magnitude above the thermal level but the clumpy behaviour occurs only over one order of magnitude.  As the electron beam propagates out into the heliosphere it spreads radially over a longer distance.  The enhanced electric field arises over 60 solar radii in length when $t=750$~s.  A large proportion of this length is at the front of the electron beam and corresponds to an increase less than one order of magnitude above the thermal level.  The enhanced density turbulence present in our model at farther distances from the Sun causes the electric field to have a clumpier distribution over the whole beam at later times.

At $t=450,750$~seconds some of the Langmuir waves are refracted out of resonance with the electron beam and can no longer be re-absorbed by the back of the electron beam, in a similar manner to what was described in Sections \ref{sec:Earth} and \ref{sec:Sun}.  The Langmuir waves that are left behind give an enhanced electric field above the background at distances behind the electron beam.  We can see this at $t=450$~seconds below 20~$R_{\odot}$ and at $t=750$~seconds below 35~$R_{\odot}$.

\subsection{Electric field distribution}

We present $P(\log E)$ for the four different times in Figure \ref{fig:LW_Pos_Time_Helio}b satisfying $\int_{E_{min}}^{E_{max}}P(\log E)d\log E=1$.  The probability distribution function was calculated using different minimum values of the electric field $E_{min}=10^{-0.5},10^{-1.0},10^{-1.5},10^{-2.0}$~mV/m at the times $t=100,200,450,750$~seconds beyond $r=4.2, 6.7, 11.6, 21.3$ solar radii, respectively.  Close to the Sun the PDF more closely resembles the unperturbed case where the PDF peaks near the highest electric fields and the lowest electric fields are produced by the front of the electron beam.  At later times the PDF becomes more evenly distributed relating to the enhanced level of density fluctuations.  The cut-off electric field means that we do not show the contribution from the front of the electron beam, particularly using $10^{-2}$~mV/m for $t=750$~seconds.  The front of the beam has a similar distribution as seen in Section \ref{sec:Earth}, increasing towards the thermal level.  The bulk shift of the PDF from high to low electric fields is due to the electron beam propagating through plasma with a decreasing mean density and hence a decreasing background electric field and Langmuir wave energy density.  We note that the distribution at $t=100$~s spans three orders of magnitude whereas the distribution at $t=750$~s spans only two orders of magnitude.  At later times the electric fields are smaller with respect to the background on account of the beam decreasing in density from the radially expanding magnetic field and propagation effects.

\subsection{Electric field moments}

\begin{figure}\center
  \includegraphics[width=0.99\columnwidth]{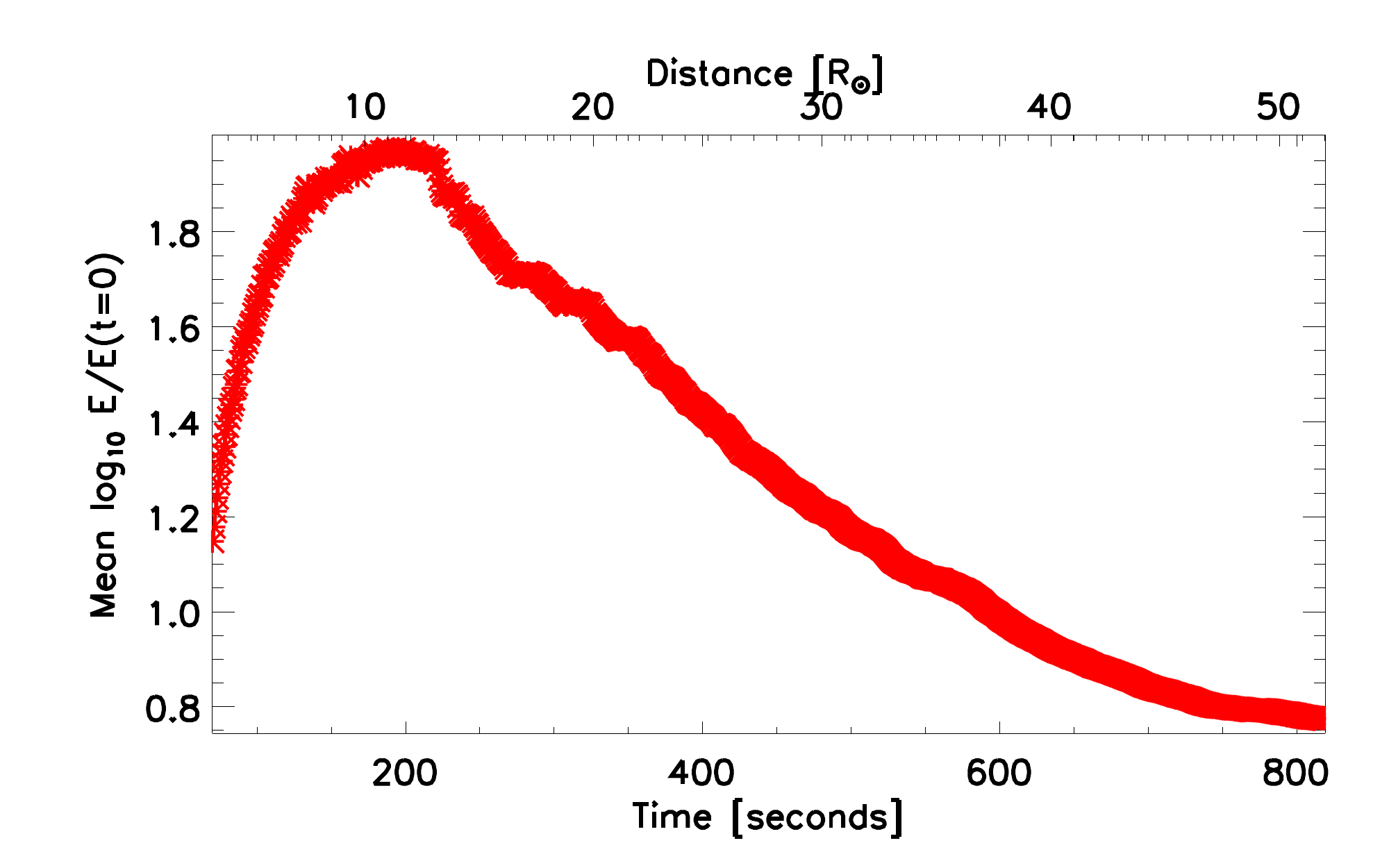}
  \includegraphics[width=0.99\columnwidth]{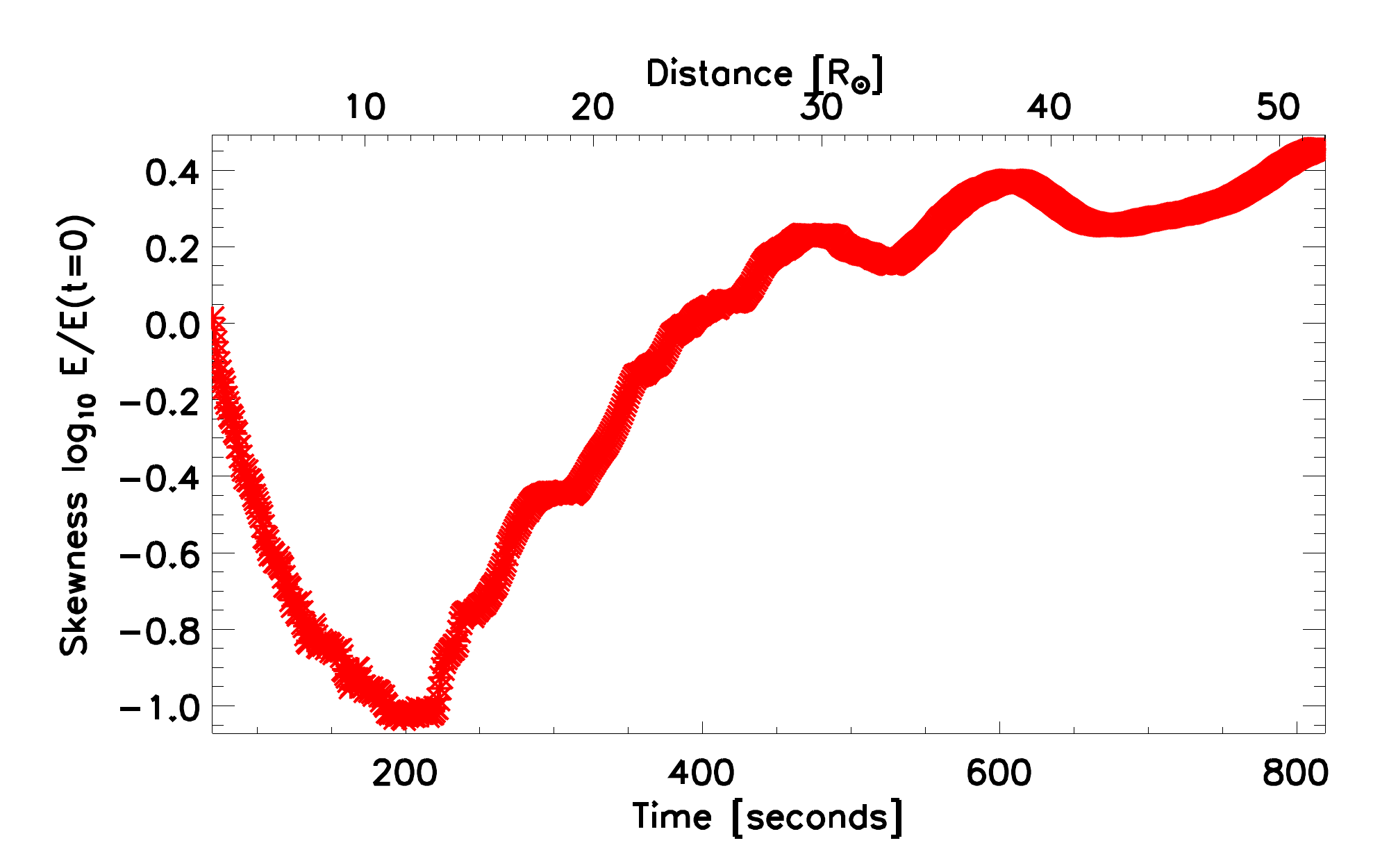}
\caption{The first three moments of $P(\log E)$ as a function of time for an electron beam injected in the solar corona and propagating out through the solar wind.  The distance is also plotted, estimated by the motion of the peak electric field in time.  The mean is normalised by the electric field produced from the thermal level of Langmuir waves.}
\label{fig:LW_Moments_Helio_Time}
\end{figure}

Figure \ref{fig:LW_Moments_Helio_Time} shows how the normalised mean and skewness of $\log E$ vary as a function of injection time.  We also show an approximation of the distance travelled by the electron beam by approximating that the peak electric field travels at a constant velocity.  The velocity was found by a straight line fit to the peak electric field as a function of time and was $v=4.5\times10^9~\rm{cm~s}^{-1}$ [$0.065~R_\odot~\rm{s}^{-1}$.  The approximation is not entirely accurate as the velocity of the peak changes as a function of time \citep[see also][]{Ratcliffe:2014aa,Krupar:2015aa} but it is a close approximation.

In a similar manner to Figure \ref{fig:LW_Moments_Sun_Time}, we initially see that the normalised mean of $\log E$ increases as a function of time together with the magnitude of the skewness.  The larger simulation box allows us to see that both moments decrease in magnitude after around 200 seconds.  The decrease in magnitude of both moments continues for the duration of the simulation.  If the beam was propagated to distances farther from the Sun we expect the normalised mean of $\log E$ would continue to decrease but the skewness of $\log E$ would reverse sign and begin to increase in magnitude in a similar way to Figure \ref{fig:LW_Moments_Earth}.  We note that the standard deviation of $\log E$ systematically decreases after around 100 seconds, representing the reduced spread of electric field values above the thermal level, enhanced by an electron beam that decreases in density as a function of time.

\section{Discussion} \label{sec:discussion}

\subsection{Resonance broadening of the beam-plasma instability}

The two main effects we found of increasing density fluctuations on the electric field distribution are a reduction in the mean of $\log E$ and a broadening of the local distribution in $\log E$, with both effects occurring at a similar rate.  We explain these effects via the process of resonance broadening due to density fluctuations following \citet{Bian:2014aa}, see also \citet{Voshchepynets:2015aa,Voshchepynets:2015ab}.  
For homogeneous plasma the wave-particle interaction has a sharp resonance function $\delta(\omega-kv)$ so electrons only interact with waves that have a phase velocity equal to their  velocity.  For inhomogeneous plasma the waves generated by the electrons are refracted over a range $\Delta v$.  Then the plasma waves averaged over density perturbation scales can be viewed resonate with electrons over a broader region in velocity space with an extent $\Delta v=\Delta \omega/k$ centred at $v=\omega/k$ \citep{Bian:2014aa}.

With a broader resonance function, the growth rate of the beam-plasma instability changes and becomes a function of the resonant width.  If the width of $\Delta v$ is small then the average slope of the electron distribution within $\Delta v$ can still be positive and waves will grow, albeit at a slightly different rate than if $\Delta v=0$.  However, if $\Delta v$ is large then the average slope can be substantially reduced or even become negative if $\Delta v$ incorporates the negative slope of the electron distribution at the highest energies or the negative slope of the background plasma at the lowest energies (see e.g. Figure 1 in \citet{Bian:2014aa}).  Resonant broadening can thus lead to weakening and a possible suppression of the beam-plasma instability.

If we approximate the wave scattering by a diffusion process then the resonance width is given by $\Delta \omega=(Dv^2)^{1/3}$  where $D$ is the diffusion constant in k-space.  For a Gaussian spectrum $D \propto (\Delta n/n)^2$ giving a resonant width of $\Delta \omega \propto (\Delta n/n)^{2/3}$.  Resonant broadening described in \citet{Bian:2014aa} focusses on the velocity dimension whilst we consider the evolution in both position and velocity.  It is not clear whether the diffusion approximation is valid in the latter scenario however; we use how the resonant width varies as a function of $\Delta n/n$ to fit the decrease in the mean and the increase in the variance of $\log E$ using
\begin{equation} \label{eqn:fit_mean}
\langle \log E \rangle = \frac{c_1}{1+c_2\left(\frac{\Delta n}{n}\right)^{c_3}},
\end{equation}
where $c_3=2/3$.  For the simulations with a constant mean background density, we found that the decrease in the mean of $\log E$ as $\Delta n/n$ increases can be fit with this distribution when we considered the entire length of the beam, the distribution above 0.01 mV/m, and the distribution of the field over a smaller subset of the beam, with different values of $c_1$ and $c_2$.

We also fit the variance of the distribution above 0.01~mV/m with the inverse of Equation \ref{eqn:fit_mean}, again with $c_3=2/3$ and found a good  match to the data.  The variance of the electric field distribution across the entire length of the beam did not change much as a function of $\Delta n/n$; both the low magnitude component from the front of the beam and high magnitude component from the bulk of the beam was always present.

The change in the resonant width as a function of $\Delta n/n$ appears to captures the behaviour of the electric field produced by a propagating electron beam.  The exact exponent $c_3$ may differ in reality but the trend of a decreasing mean field and increasing variance will likely be the same.  We considered density fluctuations only parallel to the magnetic field and so a next step would be to check whether a similar behaviour is observed for the scattering of Langmuir waves off fluctuations that are perpendicular to the direction of travel.

\subsection{Structure of the density fluctuations}

\begin{figure}\center
  \includegraphics[width=0.99\columnwidth]{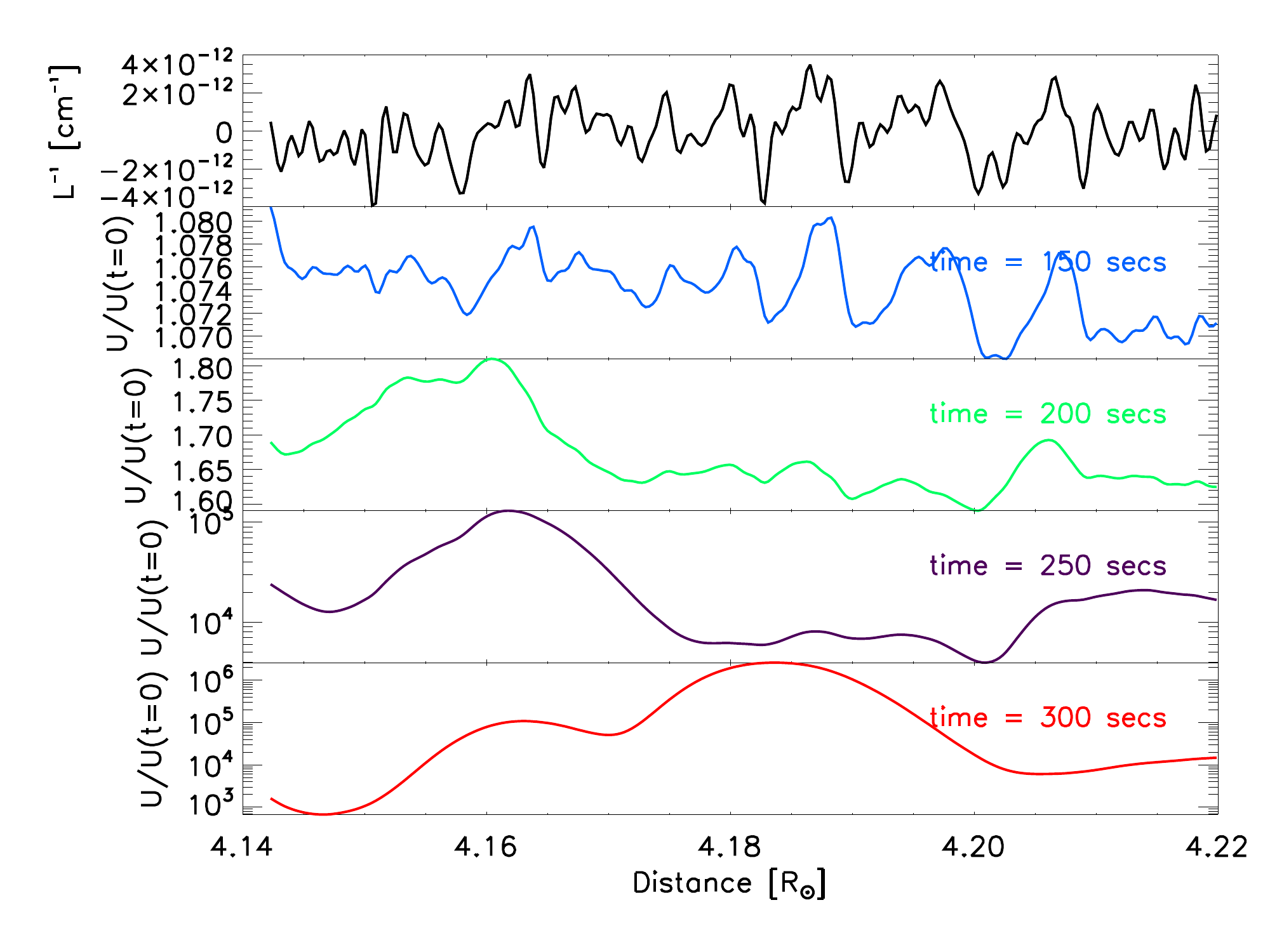}
\caption{Top: characteristic scale of the plasma inhomogeneity $L^{-1}=0.5n_e^{-1}dn_e/dx$ when $\Delta n/n=10^{-3}$ for a constant mean background electron density, over an area of space 56 Mm in length.  Bottom:  Langmuir wave energy density, normalised by the initial value, at four different times $t=150,200,250,300$~seconds.  The earliest time corresponds to spontaneous emission from the front of the beam.  The other three times correspond to wave growth from the bump-in-tail instability.  Wave propagation smooths out the fine structure present in the plasma inhomogeneity.}
\label{fig:LW_group}
\end{figure}

The magnitude of $\Delta n/n$ is not the only parameter that contributes to the effect of resonant broadening on the electric field distribution.  The length scales of the density fluctuations and the spectrum of the fluctuations play an important role.  We used length scales from $10^{10}\rightarrow10^8$~cm, within the inertial range for the solar wind \citep{Celnikier:1987aa,Chen:2013ab}, and a spectrum of $-5/3$.  The smallest length scales are the most significant for the refraction of waves because the magnitude of the spectrum is less than two.  With the same value of $\Delta n/n$ a reduction in length scales would likely increase the effect of resonant broadening on the induced electric field.  For small wavelengths outside the inertial range the spectrum steepens becoming greater than two, at which point these wavelengths likely have less of an effect for the refraction of Langmuir waves.  The spectrum of the density fluctuations may also be different close to the Sun.  For a constant value of $\Delta n/n$, decreasing the magnitude of the spectrum will increase the effect of refraction on the waves.

The structure of the Langmuir wave energy density, and hence the electric field, does not always mirror the structure of the background density fluctuations.  The propagation of waves due to the group velocity of the Langmuir waves $v_g=3v_{Te}^2/v$ smooths out the fine structure present in the background density.  Figure \ref{fig:LW_group} highlights this point by showing the energy density at four separate times together with the characteristic scale of the background plasma inhomogeneity $L^{-1}=0.5n_e^{-1}dn_e/dx$ over a small region in space 56~Mm in length.  At the earliest time, waves are due to spontaneous emission that grow (amongst other terms) proportional to $\omega_{pe}$.  The fine structure from the background electron density can be observed in the wave energy density but the magnitude above the thermal level is low.  At later times, wave growth is due to the bump-in-tail instability and the fine structure disappears as a function of time.  Given our initial conditions that the background inhomogeneity is static, the energy density will show fine structure up to a length of $d(t)v_g/v$, where $d(t)$ is the size of the electron beam at time $t$.

At $t=300$~seconds the peak in energy density has moved in space from where it was at $t=250$~seconds, at a rate equal to the group velocity around $3\times10^7~\rm{cm~s}^{-1}$.  The higher level of Langmuir waves generated by the slower, denser electrons causes a greater diffusion of electrons to lower energies.  The corresponding Langmuir waves spectrum stretches to lower phase velocities.  The high level of Langmuir waves exist longer in space before the back of the electron beam re-absorbs their energy; long enough to propagate a significant distance under their own group velocity.

\subsection{Form of the electric field distribution}

The reduction in the mean value of $\log E$ and the increase in the variance of $\log E$ from the inclusion of density fluctuations is best characterised in Figure \ref{fig:LW_SGT_Earth_PDF} where we displayed the distribution of a region of space that had high electric fields for the unperturbed case.  The exact mathematical form of the distribution is not clear.  One of the simplest expressions to compare with simulations would be a log-normal distribution \citep{Robinson:1993ab}
\begin{equation}\label{eqn:lognormal}
P(\log E) = \frac{1}{\sqrt{2\pi}\sigma_E}\exp{
\left(-\frac{(\log E - \mu_E)^2}{2\sigma_E^2}\right)}
\end{equation}
where $\mu_E$ is the mean value of $\log E$ and $\sigma_E$ is the standard deviation of $\log E$.  To compare the simulations with different $\Delta n/n$, and consequently different mean and standard deviations in $\log E$, we define a new variable $X$ such that 
\begin{equation}\label{eqn:newvar}
X = \frac{\log E - \mu_E}{\sigma_E}
\end{equation}
and compare $P(X)$ to the log-normal distribution
\begin{equation}\label{eqn:sgttest}
P(X) = \frac{1}{\sqrt{2\pi}}\exp{
\left(-\frac{X^2}{2}\right)}.
\end{equation}
To minimise the effect of the electric field varying substantially in space we analysed the PDF of the electric field obtained from the subset of the beam using the same conditions as the PDF shown in Figure \ref{fig:LW_SGT_Earth_PDF}.  To smooth the results we take the PDF of the Langmuir wave energy density over 40 seconds, from 257 to 297 seconds.

\begin{figure}\center
  \includegraphics[width=0.99\columnwidth]{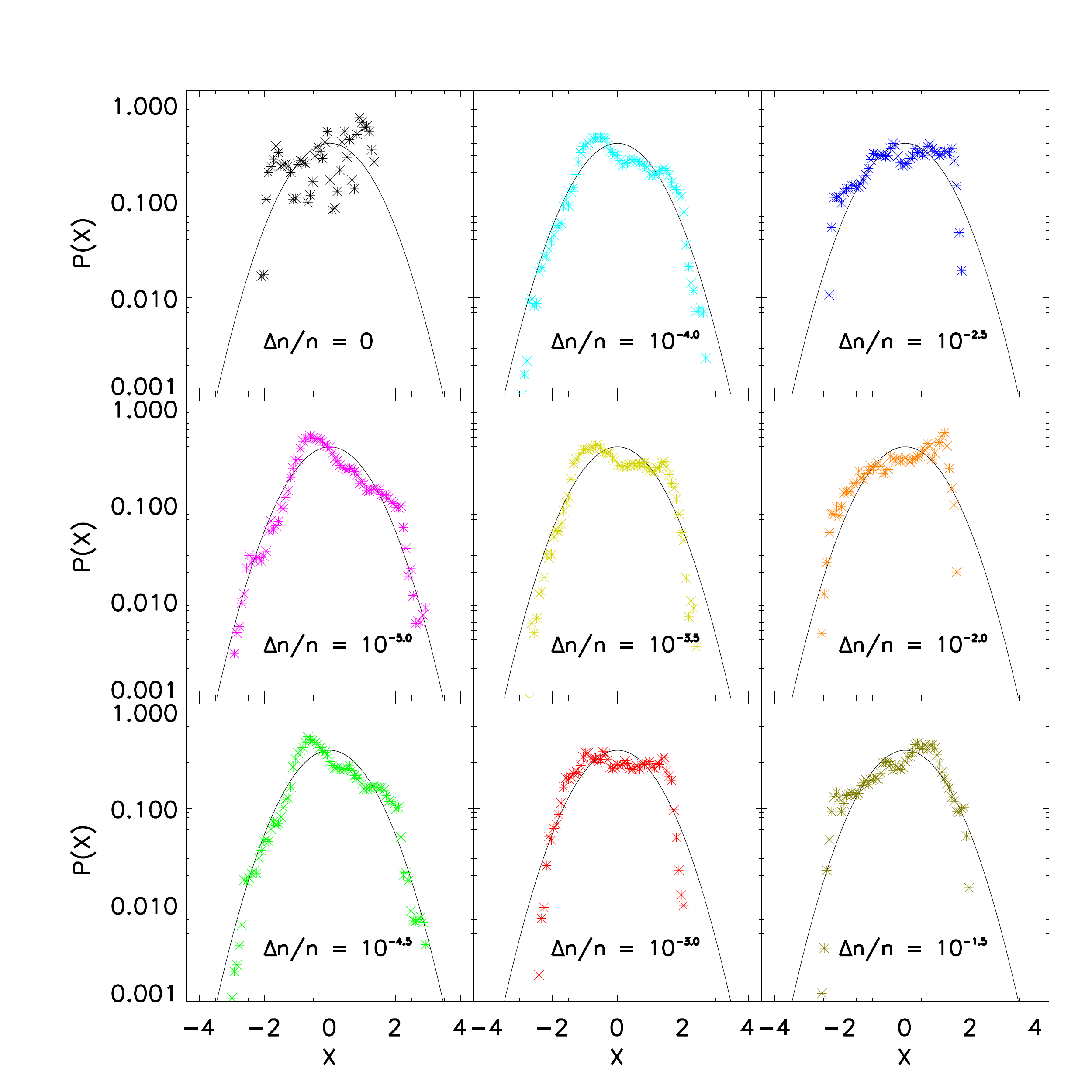}
\caption{P(X) the probability distribution function of $X = \frac{\log E - \mu_E}{\sigma_E}$ where $\mu_E,\sigma_E$ are the mean and standard deviation of $\log E$ respectively.  The simulations with different levels of $\Delta n/n$ are compared with a log-normal distribution (solid line).   The electric field distributions are sampled from the densest part of the beam and averaged over 40 seconds, similar to Figure \ref{fig:LW_SGT_Earth_PDF}.}
\label{fig:LW_SGT_Earth}
\end{figure}

Figure \ref{fig:LW_SGT_Earth} plots $P(X)$ for all nine simulations together with a curve that represents a log-normal distribution.  None of the simulations correspond particularly well to the log-normal distribution.  When $\Delta n/n>0$ there is a better correspondence with the log-normal statistics than $\Delta n/n=0$ but the distribution tends to exhibit a straight line peaked below the expected value with a negative gradient.  As $\Delta n/n$ increases the gradient of the straight line changes sign and the distribution is peaked above the expected value.  


The earlier simulations of \citet{Li:2006aa} observed log-normal distributions but used an assumption that the density fluctuations would just damp the Langmuir waves instead of the shift in k-space, an assumption that was changed in their later simulations \citep[e.g.][]{Li:2008ac}.  Under SGT the log-normal distribution occurs only under specific conditions \citep{Cairns:2007aa}.  For our simulations, the interaction between the beam-driven growth rate of waves, the refraction of waves off density fluctuations and the group velocity of waves did not produce simple log-normal distribution and expectedly depends on the sampling of space.  Further studies could be done including the reflection of waves in temporally evolving density clumps.

\subsection{Radial behaviour of the electric field distribution}

The simulations of beam transport in the solar corona and solar wind had normalised mean electric fields that initially increased and then decreased after a certain propagation time, dependent upon the magnitude of $\Delta n/n(r)$.  Higher $\Delta n/n(r)$ stifled the beam-plasma instability and caused the Langmuir waves, and hence the electric field, to decreases earlier than for a low level or no fluctuations.  It has been observed \citep[e.g.][]{Dulk:1998aa,Krupar:2014aa} that the peak type III radio intensity for interplanetary bursts occurs on average around 1~MHz.  Our simulation that travelled through the heliosphere had a normalised mean electric field that peaked around 200 seconds after beam injection.  We can see from Figure \ref{fig:LW_Pos_Time_Helio} the electric field distribution in space is centred around 12 solar radii after 200 seconds.  This corresponds in our density model to around 0.5 MHz that would create 1~MHz emission under the second harmonic.

For type III bursts the exact frequency that corresponds to the peak radio flux will be dependent upon both the electron beam properties and the background plasma properties \citep[discussed in both][]{Krupar:2014aa,Reid:2015aa}.  The expansion of the solar wind plasma, the energy density and spectrum of the electron beam, and the level of density turbulence are all important factors for the generation of Langmuir waves.  What we show is that the density turbulence can play a significant role in determining at what distance (and hence frequency) relates to the peak level of Langmuir waves.  From our simulations we see that a high level of density fluctuations suppressed Langmuir waves after only 20 seconds, well before 1~MHz plasma, and is perhaps further evidence that the solar wind close to the Sun is not as turbulent as 1~AU.

\section{Summary}

We have analysed the Langmuir wave electric field distributions generated by a propagating electron beam in the turbulent plasma of the solar corona and the solar wind.  Using weak turbulence simulations we have modelled an electron beam travelling through plasma with a varying intensity of density turbulence to observe how the fluctuations modify the distribution of the electric field.

In unperturbed plasma, the electric field distribution produced from a propagating electron beam is concentrated around the peak values and determined by the spatial profile of the beam.  The bulk of the enhanced electric field occurs in the region of space around the peak of the electron cloud. The front of the electron beam produces low-intensity electric fields on account of the low-density, high-energy electrons that populate this region.  This agrees with the absence of high electric fields observed in-situ together with the arrival of the highest energy electrons.

The presence of density fluctuations in the background plasma causes the logarithm of the electric field to become more uniformly distributed and decreases the mean field.  The effect is heightened when the intensity of the density fluctuations is increased.  We described the effect using resonance broadening approach \citep{Bian:2014aa} where electrons are able to resonate with Langmuir waves over a broader range of phase velocities on account of wave refraction off the density fluctuations.  The presence of density fluctuations naturally causes the electric field to develop a clumpy pattern, similar to what is observed in-situ by spacecraft in the solar wind.  The future missions of Solar Orbiter and Solar Probe Plus will provide a 3D view of the density close to the Sun.  Whilst our simulations used a 1D quasilinear approach, based on the angular scattering considerations in \citet{Bian:2014aa}, the average effects on Langmuir wave generation in 3D are likely to be the same but the Langmuir wave angular distribution will be different.

We found that the properties of the electric field distribution were heavily dependent on the intensity of the density turbulence and showed how the mean and the variance of PDF would change as a function of $\Delta n/n(r)$.  If density fluctuations are less pronounced close to the Sun then the upcoming missions of Solar Orbiter and Solar Probe Plus might observe electric fields to be less clumpy.  A similar variation in the electric field distribution might be present between the fast and slow solar wind if the level of turbulence is different.  We also found that the radial distance corresponding to the highest level of Langmuir waves produced above the background thermal level was dependent on the level of density fluctuations.  Under the assumption that the radio flux is proportional to the energy density of Langmuir waves, the frequency corresponding to peak flux of interplanetary type III radio bursts could give information about the local level of density turbulence in the solar wind from radio emission.

\begin{acknowledgements}
This work is supported by a STFC consolidated grant ST/L000741/1.  This work used the DiRAC Data Centric system at Durham University, operated by the Institute for Computational Cosmology on behalf of the STFC DiRAC HPC Facility (www.dirac.ac.uk. This equipment was funded by a BIS National E-infrastructure capital grant ST/K00042X/1, STFC capital grant ST/K00087X/1, DiRAC Operations grant ST/K003267/1 and Durham University. DiRAC is part of the National E-Infrastructure.  Financial support by the European Commission through the ``Radiosun'' (PEOPLE-2011-IRSES-295272) is gratefully acknowledged.  This work benefited from the Royal Society grant RG130642.
\end{acknowledgements}
\bibliographystyle{aa}
\bibliography{/Users/hamish/Documents/Papers/ubib}

\begin{thebibliography}{85}
\expandafter\ifx\csname natexlab\endcsname\relax\def\natexlab#1{#1}\fi

\bibitem[{{Bale} {et~al.}(1997){Bale}, {Burgess}, {Kellogg}, {Goetz}, \&
  {Monson}}]{Bale:1997aa}
{Bale}, S.~D., {Burgess}, D., {Kellogg}, P.~J., {Goetz}, K., \& {Monson}, S.~J.
  1997, \jgr, 102, 11281

\bibitem[{{Bian} {et~al.}(2014){Bian}, {Kontar}, \& {Ratcliffe}}]{Bian:2014aa}
{Bian}, N.~H., {Kontar}, E.~P., \& {Ratcliffe}, H. 2014, Journal of Geophysical
  Research (Space Physics), 119, 4239

\bibitem[{{Cairns} {et~al.}(2007){Cairns}, {Konkolewicz}, \&
  {Robinson}}]{Cairns:2007aa}
{Cairns}, I.~H., {Konkolewicz}, D.~L., \& {Robinson}, P.~A. 2007, Physics of
  Plasmas, 14, 042105

\bibitem[{{Cairns} \& {Robinson}(1997)}]{Cairns:1997aa}
{Cairns}, I.~H. \& {Robinson}, P.~A. 1997, \grl, 24, 369

\bibitem[{{Celnikier} {et~al.}(1983){Celnikier}, {Harvey}, {Jegou}, {Moricet},
  \& {Kemp}}]{Celnikier:1983aa}
{Celnikier}, L.~M., {Harvey}, C.~C., {Jegou}, R., {Moricet}, P., \& {Kemp}, M.
  1983, \aap, 126, 293

\bibitem[{{Celnikier} {et~al.}(1987){Celnikier}, {Muschietti}, \&
  {Goldman}}]{Celnikier:1987aa}
{Celnikier}, L.~M., {Muschietti}, L., \& {Goldman}, M.~V. 1987, \aap, 181, 138

\bibitem[{{Chen} {et~al.}(2013){Chen}, {Howes}, {Bonnell}, {Mozer}, {Klein}, \&
  {Bale}}]{Chen:2013ab}
{Chen}, C.~H.~K., {Howes}, G.~G., {Bonnell}, J.~W., {et~al.} 2013, in American
  Institute of Physics Conference Series, Vol. 1539, American Institute of
  Physics Conference Series, ed. G.~P. {Zank}, J.~{Borovsky}, R.~{Bruno},
  J.~{Cirtain}, S.~{Cranmer}, H.~{Elliott}, J.~{Giacalone}, W.~{Gonzalez},
  G.~{Li}, E.~{Marsch}, E.~{Moebius}, N.~{Pogorelov}, J.~{Spann}, \&
  O.~{Verkhoglyadova}, 143--146

\bibitem[{{Chen} {et~al.}(2012){Chen}, {Mallet}, {Schekochihin}, {Horbury},
  {Wicks}, \& {Bale}}]{Chen:2012aa}
{Chen}, C.~H.~K., {Mallet}, A., {Schekochihin}, A.~A., {et~al.} 2012, \apj,
  758, 120

\bibitem[{{Coste} {et~al.}(1975){Coste}, {Reinisch}, {Montes}, \&
  {Silevitch}}]{Coste:1975aa}
{Coste}, J., {Reinisch}, G., {Montes}, C., \& {Silevitch}, M.~B. 1975, Physics
  of Fluids, 18, 679

\bibitem[{{Daldorff} {et~al.}(2011){Daldorff}, {P{\'e}cseli}, {Trulsen},
  {Ulriksen}, {Eliasson}, \& {Stenflo}}]{Daldorff:2011aa}
{Daldorff}, L.~K.~S., {P{\'e}cseli}, H.~L., {Trulsen}, J.~K., {et~al.} 2011,
  Physics of Plasmas, 18, 052107

\bibitem[{{Drummond} \& {Pines}(1964)}]{Drummond:1964aa}
{Drummond}, W.~E. \& {Pines}, D. 1964, Annals of Physics, 28, 478

\bibitem[{{Dulk} {et~al.}(1998){Dulk}, {Leblanc}, {Robinson}, {Bougeret}, \&
  {Lin}}]{Dulk:1998aa}
{Dulk}, G.~A., {Leblanc}, Y., {Robinson}, P.~A., {Bougeret}, J.-L., \& {Lin},
  R.~P. 1998, \jgr, 103, 17223

\bibitem[{{Foroutan} {et~al.}(2007){Foroutan}, {Robinson}, {Sobhanian},
  {Moslehi-Fard}, {Li}, \& {Cairns}}]{Foroutan:2007aa}
{Foroutan}, G.~R., {Robinson}, P.~A., {Sobhanian}, S., {et~al.} 2007, Physics
  of Plasmas, 14, 012903

\bibitem[{{Ginzburg} \& {Zhelezniakov}(1958)}]{Ginzburg:1958aa}
{Ginzburg}, V.~L. \& {Zhelezniakov}, V.~V. 1958, \sovast, 2, 653

\bibitem[{{Grognard}(1985)}]{Grognard:1985aa}
{Grognard}, R.~J.-M. 1985, {Propagation of electron streams}, ed. D.~J.
  {McLean} \& N.~R. {Labrum}, 253--286

\bibitem[{{Gurnett} \& {Anderson}(1976)}]{Gurnett:1976aa}
{Gurnett}, D.~A. \& {Anderson}, R.~R. 1976, Science, 194, 1159

\bibitem[{{Gurnett} \& {Anderson}(1977)}]{Gurnett:1977aa}
---. 1977, \jgr, 82, 632

\bibitem[{{Gurnett} {et~al.}(1978){Gurnett}, {Anderson}, {Scarf}, \&
  {Kurth}}]{Gurnett:1978aa}
{Gurnett}, D.~A., {Anderson}, R.~R., {Scarf}, F.~L., \& {Kurth}, W.~S. 1978,
  \jgr, 83, 4147

\bibitem[{{Gurnett} {et~al.}(1980){Gurnett}, {Anderson}, \&
  {Tokar}}]{Gurnett:1980aa}
{Gurnett}, D.~A., {Anderson}, R.~R., \& {Tokar}, R.~L. 1980, in IAU Symposium,
  Vol.~86, Radio Physics of the Sun, ed. M.~R. {Kundu} \& T.~E. {Gergely},
  369--378

\bibitem[{{Gurnett} \& {Frank}(1975)}]{Gurnett:1975aa}
{Gurnett}, D.~A. \& {Frank}, L.~A. 1975, \solphys, 45, 477

\bibitem[{{Gurnett} {et~al.}(1993){Gurnett}, {Hospodarsky}, {Kurth},
  {Williams}, \& {Bolton}}]{Gurnett:1993aa}
{Gurnett}, D.~A., {Hospodarsky}, G.~B., {Kurth}, W.~S., {Williams}, D.~J., \&
  {Bolton}, S.~J. 1993, \jgr, 98, 5631

\bibitem[{{Hannah} {et~al.}(2009){Hannah}, {Kontar}, \&
  {Sirenko}}]{Hannah:2009aa}
{Hannah}, I.~G., {Kontar}, E.~P., \& {Sirenko}, O.~K. 2009, \apjl, 707, L45

\bibitem[{{Henri} {et~al.}(2010){Henri}, {Califano}, {Briand}, \&
  {Mangeney}}]{Henri:2010aa}
{Henri}, P., {Califano}, F., {Briand}, C., \& {Mangeney}, A. 2010, Journal of
  Geophysical Research (Space Physics), 115, A06106

\bibitem[{{Holman} {et~al.}(2011){Holman}, {Aschwanden}, {Aurass}, {Battaglia},
  {Grigis}, {Kontar}, {Liu}, {Saint-Hilaire}, \& {Zharkova}}]{Holman:2011aa}
{Holman}, G.~D., {Aschwanden}, M.~J., {Aurass}, H., {et~al.} 2011, \ssr, 159,
  107

\bibitem[{{Howes} {et~al.}(2012){Howes}, {Bale}, {Klein}, {Chen}, {Salem}, \&
  {TenBarge}}]{Howes:2012aa}
{Howes}, G.~G., {Bale}, S.~D., {Klein}, K.~G., {et~al.} 2012, \apjl, 753, L19

\bibitem[{{Karlick{\'y}} \& {Kontar}(2012)}]{Karlicky:2012aa}
{Karlick{\'y}}, M. \& {Kontar}, E.~P. 2012, \aap, 544, A148

\bibitem[{{Kontar}(2001{\natexlab{a}})}]{Kontar:2001ab}
{Kontar}, E.~P. 2001{\natexlab{a}}, \solphys, 202, 131

\bibitem[{{Kontar}(2001{\natexlab{b}})}]{Kontar:2001aa}
---. 2001{\natexlab{b}}, \aap, 375, 629

\bibitem[{{Kontar}(2001{\natexlab{c}})}]{Kontar:2001ad}
---. 2001{\natexlab{c}}, Computer Physics Communications, 138, 222

\bibitem[{{Kontar} \& {Reid}(2009)}]{Kontar:2009aa}
{Kontar}, E.~P. \& {Reid}, H.~A.~S. 2009, \apjl, 695, L140

\bibitem[{{Krafft} {et~al.}(2013){Krafft}, {Volokitin}, \&
  {Krasnoselskikh}}]{Krafft:2013aa}
{Krafft}, C., {Volokitin}, A.~S., \& {Krasnoselskikh}, V.~V. 2013, \apj, 778,
  111

\bibitem[{{Krafft} {et~al.}(2014){Krafft}, {Volokitin}, {Krasnoselskikh}, \&
  {de Wit}}]{Krafft:2014ab}
{Krafft}, C., {Volokitin}, A.~S., {Krasnoselskikh}, V.~V., \& {de Wit}, T.~D.
  2014, Journal of Geophysical Research (Space Physics), 119, 9369

\bibitem[{{Krucker} {et~al.}(2007){Krucker}, {Kontar}, {Christe}, \&
  {Lin}}]{Krucker:2007aa}
{Krucker}, S., {Kontar}, E.~P., {Christe}, S., \& {Lin}, R.~P. 2007, \apjl,
  663, L109

\bibitem[{{Krucker} {et~al.}(2009){Krucker}, {Oakley}, \&
  {Lin}}]{Krucker:2009aa}
{Krucker}, S., {Oakley}, P.~H., \& {Lin}, R.~P. 2009, \apj, 691, 806

\bibitem[{{Krupar} {et~al.}(2015){Krupar}, {Kontar}, {Soucek}, {Santolik},
  {Maksimovic}, \& {Kruparova}}]{Krupar:2015aa}
{Krupar}, V., {Kontar}, E.~P., {Soucek}, J., {et~al.} 2015, \aap, 580, A137

\bibitem[{{Krupar} {et~al.}(2014){Krupar}, {Maksimovic}, {Santolik}, {Kontar},
  {Cecconi}, {Hoang}, {Kruparova}, {Soucek}, {Reid}, \&
  {Zaslavsky}}]{Krupar:2014aa}
{Krupar}, V., {Maksimovic}, M., {Santolik}, O., {et~al.} 2014, \solphys, 289,
  3121

\bibitem[{{Leblanc} {et~al.}(1998){Leblanc}, {Dulk}, \&
  {Bougeret}}]{Leblanc:1998aa}
{Leblanc}, Y., {Dulk}, G.~A., \& {Bougeret}, J.-L. 1998, \solphys, 183, 165

\bibitem[{{Li} \& {Cairns}(2014)}]{Li:2014aa}
{Li}, B. \& {Cairns}, I.~H. 2014, \solphys, 289, 951

\bibitem[{{Li} {et~al.}(2008){Li}, {Cairns}, \& {Robinson}}]{Li:2008ac}
{Li}, B., {Cairns}, I.~H., \& {Robinson}, P.~A. 2008, Journal of Geophysical
  Research (Space Physics), 113, 6104

\bibitem[{{Li} {et~al.}(2012){Li}, {Cairns}, \& {Robinson}}]{Li:2012aa}
---. 2012, \solphys, 279, 173

\bibitem[{{Li} {et~al.}(2006){Li}, {Robinson}, \& {Cairns}}]{Li:2006aa}
{Li}, B., {Robinson}, P.~A., \& {Cairns}, I.~H. 2006, Physics of Plasmas, 13,
  082305

\bibitem[{{Lin} {et~al.}(1981){Lin}, {Potter}, {Gurnett}, \&
  {Scarf}}]{Lin:1981aa}
{Lin}, R.~P., {Potter}, D.~W., {Gurnett}, D.~A., \& {Scarf}, F.~L. 1981, \apj,
  251, 364

\bibitem[{{Magelssen} \& {Smith}(1977)}]{Magelssen:1977aa}
{Magelssen}, G.~R. \& {Smith}, D.~F. 1977, \solphys, 55, 211

\bibitem[{{Maksimovic} {et~al.}(2005){Maksimovic}, {Zouganelis}, {Chaufray},
  {Issautier}, {Scime}, {Littleton}, {Marsch}, {McComas}, {Salem}, {Lin}, \&
  {Elliott}}]{Maksimovic:2005aa}
{Maksimovic}, M., {Zouganelis}, I., {Chaufray}, J.-Y., {et~al.} 2005, Journal
  of Geophysical Research (Space Physics), 110, A09104

\bibitem[{{Malaspina} {et~al.}(2010){Malaspina}, {Kellogg}, {Bale}, \&
  {Ergun}}]{Malaspina:2010ab}
{Malaspina}, D.~M., {Kellogg}, P.~J., {Bale}, S.~D., \& {Ergun}, R.~E. 2010,
  \apj, 711, 322

\bibitem[{{Mann} {et~al.}(1999){Mann}, {Jansen}, {MacDowall}, {Kaiser}, \&
  {Stone}}]{Mann:1999aa}
{Mann}, G., {Jansen}, F., {MacDowall}, R.~J., {Kaiser}, M.~L., \& {Stone},
  R.~G. 1999, \aap, 348, 614

\bibitem[{{Marsch} \& {Tu}(1990)}]{Marsch:1990aa}
{Marsch}, E. \& {Tu}, C.-Y. 1990, \jgr, 95, 11945

\bibitem[{{Mel'nik} {et~al.}(2000){Mel'nik}, {Kontar}, \&
  {Lapshin}}]{Melnik:2000aa}
{Mel'nik}, V.~N., {Kontar}, E.~P., \& {Lapshin}, V.~I. 2000, \solphys, 196, 199

\bibitem[{{Mel'Nik} {et~al.}(1999){Mel'Nik}, {Lapshin}, \&
  {Kontar}}]{MelNik:1999ab}
{Mel'Nik}, V.~N., {Lapshin}, V., \& {Kontar}, E. 1999, \solphys, 184, 353

\bibitem[{{Melrose}(1980)}]{Melrose:1980aa}
{Melrose}, D.~B. 1980, {Plasma astrohysics. Nonthermal processes in diffuse
  magnetized plasmas - Vol.1: The emission, absorption and transfer of waves in
  plasmas; Vol.2: Astrophysical applications}

\bibitem[{{Melrose} {et~al.}(1986){Melrose}, {Cairns}, \&
  {Dulk}}]{Melrose:1986aa}
{Melrose}, D.~B., {Cairns}, I.~H., \& {Dulk}, G.~A. 1986, \aap, 163, 229

\bibitem[{{Meyer-Vernet}(1979)}]{Meyer-Vernet:1979aa}
{Meyer-Vernet}, N. 1979, \jgr, 84, 5373

\bibitem[{{Muschietti} {et~al.}(1985){Muschietti}, {Goldman}, \&
  {Newman}}]{Muschietti:1985aa}
{Muschietti}, L., {Goldman}, M.~V., \& {Newman}, D. 1985, \solphys, 96, 181

\bibitem[{{Newkirk}(1961)}]{Newkirk:1961aa}
{Newkirk}, Jr., G. 1961, \apj, 133, 983

\bibitem[{{Parker}(1958)}]{Parker:1958aa}
{Parker}, E.~N. 1958, \apj, 128, 664

\bibitem[{{P{\'e}cseli} \& {P{\'e}cseli}(2014)}]{Pecseli:2014aa}
{P{\'e}cseli}, H.~L. \& {P{\'e}cseli}. 2014, Journal of Plasma Physics, 80, 745

\bibitem[{{P{\'e}cseli} \& {Trulsen}(1992)}]{Pecseli:1992aa}
{P{\'e}cseli}, H.~L. \& {Trulsen}, J. 1992, \physscr, 46, 159

\bibitem[{{Ratcliffe} {et~al.}(2012){Ratcliffe}, {Bian}, \&
  {Kontar}}]{Ratcliffe:2012aa}
{Ratcliffe}, H., {Bian}, N.~H., \& {Kontar}, E.~P. 2012, \apj, 761, 176

\bibitem[{{Ratcliffe} {et~al.}(2014){Ratcliffe}, {Kontar}, \&
  {Reid}}]{Ratcliffe:2014aa}
{Ratcliffe}, H., {Kontar}, E.~P., \& {Reid}, H.~A.~S. 2014, \aap, 572, A111

\bibitem[{{Reid} \& {Kontar}(2010)}]{Reid:2010aa}
{Reid}, H.~A.~S. \& {Kontar}, E.~P. 2010, \apj, 721, 864

\bibitem[{{Reid} \& {Kontar}(2013)}]{Reid:2013aa}
---. 2013, \solphys, 285, 217

\bibitem[{{Reid} \& {Kontar}(2015)}]{Reid:2015aa}
---. 2015, \aap, 577, A124

\bibitem[{{Reid} {et~al.}(2014){Reid}, {Vilmer}, \& {Kontar}}]{Reid:2014aa}
{Reid}, H.~A.~S., {Vilmer}, N., \& {Kontar}, E.~P. 2014, \aap, 567, A85

\bibitem[{{Robinson}(1992)}]{Robinson:1992aa}
{Robinson}, P.~A. 1992, \solphys, 139, 147

\bibitem[{{Robinson} \& {Cairns}(1993)}]{Robinson:1993ab}
{Robinson}, P.~A. \& {Cairns}, I.~H. 1993, \apj, 418, 506

\bibitem[{{Robinson} {et~al.}(1993){Robinson}, {Cairns}, \&
  {Gurnett}}]{Robinson:1993aa}
{Robinson}, P.~A., {Cairns}, I.~H., \& {Gurnett}, D.~A. 1993, \apj, 407, 790

\bibitem[{{Robinson} {et~al.}(2004){Robinson}, {Li}, \&
  {Cairns}}]{Robinson:2004aa}
{Robinson}, P.~A., {Li}, B., \& {Cairns}, I.~H. 2004, Physical Review Letters,
  93, 235003

\bibitem[{{Ryutov}(1969)}]{Ryutov:1969aa}
{Ryutov}, D.~D. 1969, Soviet Journal of Experimental and Theoretical Physics,
  30, 131

\bibitem[{{Sigsbee} {et~al.}(2004){Sigsbee}, {Kletzing}, {Gurnett}, {Pickett},
  {Balogh}, \& {Lucek}}]{Sigsbee:2004aa}
{Sigsbee}, K., {Kletzing}, C., {Gurnett}, D., {et~al.} 2004, Annales
  Geophysicae, 22, 2337

\bibitem[{{Sittler} \& {Guhathakurta}(1999)}]{Sittler:1999aa}
{Sittler}, Jr., E.~C. \& {Guhathakurta}, M. 1999, \apj, 523, 812

\bibitem[{{Smith} \& {Sime}(1979)}]{Smith:1979aa}
{Smith}, D.~F. \& {Sime}, D. 1979, \apj, 233, 998

\bibitem[{{Sturrock}(1964)}]{Sturrock:1964aa}
{Sturrock}, P.~A. 1964, NASA Special Publication, 50, 357

\bibitem[{{Takakura} \& {Shibahashi}(1976)}]{Takakura:1976aa}
{Takakura}, T. \& {Shibahashi}, H. 1976, \solphys, 46, 323

\bibitem[{{Tsiklauri}(2011)}]{Tsiklauri:2011aa}
{Tsiklauri}, D. 2011, Physics of Plasmas, 18, 052903

\bibitem[{{Umeda}(2007)}]{Umeda:2007aa}
{Umeda}, T. 2007, Nonlinear Processes in Geophysics, 14, 671

\bibitem[{{Vedenov}(1963)}]{Vedenov:1963aa}
{Vedenov}, A.~A. 1963, Journal of Nuclear Energy, 5, 169

\bibitem[{{Vidojevic} {et~al.}(2012){Vidojevic}, {Zaslavsky}, {Maksimovic},
  {Atanackovic}, {Hoang}, \& {Drazic}}]{Vidojevic:2012aa}
{Vidojevic}, S., {Zaslavsky}, A., {Maksimovic}, M., {et~al.} 2012, Publications
  of the Astronomical Society ''Rudjer Boskovic'', 11, 343

\bibitem[{{Voshchepynets} \& {Krasnoselskikh}(2015)}]{Voshchepynets:2015ab}
{Voshchepynets}, A. \& {Krasnoselskikh}, V. 2015, Journal of Geophysical
  Research (Space Physics), 120, 10

\bibitem[{{Voshchepynets} {et~al.}(2015){Voshchepynets}, {Krasnoselskikh},
  {Artemyev}, \& {Volokitin}}]{Voshchepynets:2015aa}
{Voshchepynets}, A., {Krasnoselskikh}, V., {Artemyev}, A., \& {Volokitin}, A.
  2015, \apj, 807, 38

\bibitem[{{Woo}(1996)}]{Woo:1996aa}
{Woo}, R. 1996, \apss, 243, 97

\bibitem[{{Woo} {et~al.}(1995){Woo}, {Armstrong}, {Bird}, \&
  {Patzold}}]{Woo:1995aa}
{Woo}, R., {Armstrong}, J.~W., {Bird}, M.~K., \& {Patzold}, M. 1995, \grl, 22,
  329

\bibitem[{{Zaitsev} {et~al.}(1972){Zaitsev}, {Mityakov}, \&
  {Rapoport}}]{Zaitsev:1972aa}
{Zaitsev}, V.~V., {Mityakov}, N.~A., \& {Rapoport}, V.~O. 1972, \solphys, 24,
  444

\bibitem[{{Zaslavsky} {et~al.}(2010){Zaslavsky}, {Volokitin}, {Krasnoselskikh},
  {Maksimovic}, \& {Bale}}]{Zaslavsky:2010aa}
{Zaslavsky}, A., {Volokitin}, A.~S., {Krasnoselskikh}, V.~V., {Maksimovic}, M.,
  \& {Bale}, S.~D. 2010, Journal of Geophysical Research (Space Physics), 115,
  A08103

\bibitem[{{Zheleznyakov} \& {Zaitsev}(1970)}]{Zheleznyakov:1970aa}
{Zheleznyakov}, V.~V. \& {Zaitsev}, V.~V. 1970, \sovast, 14, 47

\bibitem[{{Ziebell} {et~al.}(2015){Ziebell}, {Yoon}, {Petruzzellis}, {Gaelzer},
  \& {Pavan}}]{Ziebell:2015aa}
{Ziebell}, L.~F., {Yoon}, P.~H., {Petruzzellis}, L.~T., {Gaelzer}, R., \&
  {Pavan}, J. 2015, \apj, 806, 237

\end{thebibliography}

\end{document}